\documentclass[twocolumn,showpacs,
  aps,superscriptaddress,
  prd,notitlepage,showkeys,
  nofootinbib,floatfix]{revtex4-1}

\usepackage[normalem]{ulem}
\usepackage{amssymb}
\usepackage{amsmath}
\usepackage{graphicx}
\usepackage{dcolumn}
\usepackage[colorlinks,urlcolor=blue,citecolor=blue,linkcolor=blue]{hyperref}
\usepackage{color,units}
\usepackage[dvipsnames]{xcolor} 
\usepackage{lineno}
\usepackage{xspace}
\usepackage{longtable} 
\usepackage{float}  
 \usepackage{aas_macros}
 \usepackage{amsfonts,wasysym,epsfig,verbatim,subfigure,bm,mathrsfs,lipsum}
\usepackage{comment}

\newcommand{\norm}[1]{\left\lVert#1\right\rVert}
\newcommand{\mat}[1]{\mathbf{#1}}

\newcommand{\RNum}[1]{\uppercase\expandafter{\romannumeral #1\relax}}

\begin{document}

\newcommand{\IUCAA}{Inter-University Centre for Astronomy and
  Astrophysics, Post Bag 4, Ganeshkhind, Pune 411 007, India}

\newcommand{\IITK}{Department of Physics, Indian Institute of Technology, Kanpur 208016, India}

\newcommand{\NIKHEF}{Nikhef - National Institute for Subatomic Physics, Science Park, 1098 XG Amsterdam, Netherlands}

\newcommand{\UVA}{Institute for High-Energy Physics, University of Amsterdam, Science Park, 1098 XG Amsterdam, Netherlands}

\newcommand{\WSU}{Department of Physics and Astronomy, Washington State University, 1245 Webster, Pullman, Washington 99164-2814, U.S.A.}

\newcommand{\GRAPS}{Institute for Gravitational and Subatomic Physics, Utrecht University, Princetonplein 1, 3584 CC Utrecht, Netherlands}

\title{Recognizing black holes in gravitational-wave observations: Challenges in telling apart impostors in mass-gap
binaries}

\author{Sayak Datta}
\affiliation{\IUCAA}
\email{skdatta@iucaa.in}

\author{Khun Sang Phukon}
\affiliation{\NIKHEF}
\affiliation{\UVA}
\affiliation{\GRAPS}
\email{k.s.phukon@nikhef.nl}

\author{Sukanta Bose}
\affiliation{\IUCAA}
\affiliation{\WSU}
\email{sukanta@iucaa.in}


\begin{abstract}
We study how by careful monitoring of the presence or absence of tidal deformability (TD) and tidal-heating (TH) in the inspiral signal of compact object binaries in ground-based gravitational wave (GW) detectors, one can test if its components are black holes or not.
The former property (TD) is finite for neutron stars but vanishes for black holes (in general relativity), whereas the latter is finite for black holes and negligible for neutron stars, and 
affects the GW phase evolution of binaries in a distinctly different way. 
We introduce waveform parameters that characterize the strength of tidal-heating, and are zero when there is no horizon. We develop Bayesian methods that use TD and TH for distinguishing the presence or absence of horizons in a binary. 
This is timely owing to several claims that these stellar-mass objects, especially, with masses heavier than those of neutron stars, may not have a horizon but may be black hole mimickers or exotic compact objects (ECOs).  It is also astrophysically important to have the tools to test the presence or absence of horizons in mass-gap binaries and, thereby, help detect the heaviest neutron star or the lightest black hole. A proper accounting of tidal-heating in binary waveform models will also be critical for an unbiased measurement of characteristics of the equation of state of neutron stars in GW observations of binaries containing them -- or even to probe the existence of ECOs. We show that purely based on GW waveforms it will not be possible to discern binary horizons in the mass gap in Advanced LIGO, Virgo and KAGRA detectors unless the binary is within a few tens of Mpc. However, third generation ground-based detectors will be able to do so for binaries a few hundred Mpc away.

\end{abstract}

\maketitle
\section{Introduction}
\label{intro}

In recent times, the discovery by 
LIGO-Virgo detectors of several compact binary coalescences (CBCs)  has ushered in the era of gravitational wave (GW) astronomy~\cite{LIGOScientific:2018mvr,LIGOScientific:2020ibl}. The LIGO-Virgo Collaboration also observed the binary neutron (BNS) star merger GW170817~\cite{gw170817}. 
These observations provided a fillip to tests of general relativity (GR) in the strong-field regime~\cite{LIGOScientific:2019fpa,Abbott:2018lct};
e.g., stringent bounds on the mass of the graviton and violations of Lorentz invariance have been placed~\cite{TheLIGOScientific:2016pea, TheLIGOScientific:2016src, Abbott:2017vtc}. 
Significantly, it has also become possible to test the nature of the compact objects in binaries. 
The deduced compactness of the components 
has led to the conclusion that they are either black holes (BHs) or neutron stars (NSs). 
In the case of GW170817, radius measurements were made~\cite{Abbott:2018exr,De:2018uhw} that strongly disfavor them as BHs. 
A similar claim may be posited for the other BNS contender GW190425~\cite{Abbott:2020uma}. However, for the other LIGO-Virgo binaries (which are much heavier than GW170817 or GW190425)~\cite{LIGOScientific:2018mvr}, it remains to be conclusively proven that their components are indeed BHs of GR and not, say, some exotic compact objects (ECOs)~\cite{Yunes:2016jcc,Cardoso:2016oxy,Aneesh:2018hlp}. 

If the binaries show up with measured components masses 
in the mass-gap $\sim 2-5 M_\odot$~\cite{Fryer_2012}, then it poses the immediate challenge of determining whether the component(s) with mass(es) in the gap are NSs or BHs. Either occurrence will be significant, for it will either raise the maximum known mass of a NS or lower the minimum known mass of a BH. These issues make it imperative that  methods be devised to discern compact  objects with horizon from those without. In this work we study if the presence of horizon can be detected in  binaries in the mass-gap by LIGO-Virgo. We also include in our study the mass range $1-2 M_\odot$ where neutron-star masses commonly occur.

Apart from NS and BH, ECOs may also occur in the same mass range.
Multiple models of ECOs have been proposed. These include Planck-scale modifications of BH horizons~\cite{Lunin:2001jy, Almheiri:2012rt}, gravastars,~\cite{Mazur:2004fk}, and
boson stars~\cite{Liebling:2012fv} -- to name a few.
In light of such proposals,
it becomes necessary to devise
strategies to tell them apart from BHs. 
Several tests have been proposed to probe the black-holeness
of the compact objects in a binary.
Distinguishing binary merger remnants from BHs in the postmerger phase using {\it echoes} has initiated rigorous modelling and search for those features in GW data~\cite{Cardoso:2016rao, Cardoso:2016oxy, Maggio:2019zyv, Tsang:2019zra,Abedi:2016hgu, Westerweck:2017hus, Cardoso:2019rvt, Chen:2020htz, Xin:2021zir}. Measurement of tidal deformability (TD)~\cite{Cardoso:2017cfl, Sennett:2017etc, Maselli:2017cmm, Datta:2021hvm} and spin-induced multipole moments \cite{Krishnendu:2017shb, Datta:2019euh, Bianchi:2020bxa, Mukherjee:2020how, Datta:2020axm} from the late inspiral can also be used to test black-holeness. 
In this paper, we expand on past work to study how difficult it is to perform a horizon test by using GWs emitted during the inspiral phase of binary coalescences. For this purpose we include terms in the binary waveform phase beyond the point-particle ones that arise due to the material characteristics of the objects or the presence of horizons. In particular, we introduce two new best measured horizon parameters for stellar-mass binaries in ground-based detectors. Their precise measurement in binary observations is useful in probing the existence of horizons in those systems.

Owing to their causal structure, BHs in GR are perfect absorbers that behave as  dissipative systems~\cite{MembraneParadigm,Damour_viscous,Poisson:2009di,Cardoso:2012zn}. 
The defining feature of a BH is the presence of its horizon, which is a null surface and a one-way membrane. It is due to the presence of the horizon that a BH in a binary absorbs energy and angular momentum from the orbit. This phenomenon is called tidal-heating (TH)~\cite{Hartle:1973zz,Hughes:2001jr,PoissonWill}. Energy loss via  TH backreacts on the binary's evolution, resulting in a shift in the phase of the GWs emitted by the system. Therefore, the absence of a horizon -- or any kind of change in the near horizon structure that modifies this absorption -- will leave its imprint in the phasing of GWs emitted. A careful observation thus has the potential to measure these differences in the GW phase. 

Indeed, TH has been proposed for probing the presence of horizons along with the existence of higher dimensions and quantum effects at horizon scale \cite{Chakraborty:2021gdf, Datta:2020rvo, Agullo:2020hxe, Datta:2021row, Sago:2021iku}. 
Its importance in identifying horizons of intermediate-mass and supermassive compact objects has been examined for the space-mission LISA~\cite{Datta:2019euh, Maselli:2017cmm, Datta:2019epe}. In the current work, we study its usefulness for stellar mass binaries -- of the type observable by ground-based GW detectors like  LIGO, Virgo, and KAGRA \cite{Akutsu:2020zlw}.

The TH of a black hole or any other star can be expressed in similar mathematical forms if the viscosity coefficient $(\eta)$ of a BH is identified with its mass~\cite{Glampedakis:2013jya}. For NS, one has  $\eta^{}_{\rm NS} \sim 10^4 \left(\frac{\rho}{10^{14} \textrm{gm}- \textrm{cm}^{-3}}\right)^{5/4} \left(\frac{10^8 \textrm{K}}{T}\right)^2 \textrm{cm}^2 \textrm{s}^{-1}$, and for a BH, its form is $\eta^{}_{\rm BH} \sim 8.6 \times 10^{14} \left(\frac{M}{M_{\odot}}\right) \textrm{cm}^2 \textrm{s}^{-1}$. 
Since the correction in GW phase due to TH is proportional to $\eta$, for an NS that correction is 10 orders of magnitude smaller than BH~\cite{Glampedakis:2013jya}. While this distinction presents an interesting prospect for observational exploitation, as we show here the magnitude of TH for binary black holes (BBHs) remains small and is useful for discerning the presence of horizons for very large $M$ (as for EMRI central objects in LISA) or for strong signals. This implies that for  stellar-mass BHs, detection of TH in
LIGO-Virgo will require
the binary to be within tens of Mpc, as shown below. While the occurrence of such a golden binary is not impossible,  GW observations to date rule it as improbable. Nevertheless, for completeness of TH analysis, we examine this case in this work. For more realistic BBH distances, a detection of TH and its utilization for discerning horizons will have to wait for third generation detectors. The formalism initiated here for accounting for TH will be relevant for 
those detectors as well. 

Another property of compact objects that leaves an imprint on GWs is tidal {\em deformability} (TD). 
A body immersed in an external tidal field, such as due to a binary companion, experiences an
induced quadrupole moment.
That moment 
is proportional to 
the tidal field, and the proportionality factor is the tidal deformability $(\lambda)$.  This tidal deformation in turn affects the binary's orbital motion and the emitted GWs. The GW phasing carries an imprint of the dimensionless tidal deformability $(\Lambda_i \equiv \lambda_i/m_i^5)$ of the two masses $m_{i=1,2}$~ \cite{Flanagan:2007ix}. 
Material bodies, such as NS, have substantial $\Lambda_i$ values~\cite{Abbott:2018exr,Abbott:2020uma},
but black holes have a vanishing value~\cite{Damour:2009vw,Binnington:2009bb}.\footnote{See, however, Ref.~\cite{Chakravarti:2018vlt} for an example of a non-GR result, and Ref. \cite{Brustein:2020tpg} for quantum BHs.}
Hence, using appropriate modeling it is possible to measure $\Lambda_i$ and probe the properties of the bodies. More than TH, it is TD that we find to have a dominating influence in recognizing the absence of horizons in a stellar-mass binary, particularly, when the components masses are around $1-2 M_\odot$. TD decreases with mass, and above this range is vanishingly small for realistic neutron star equations of state.

Mass-gap objects can be as heavy as $\sim 5M_{\odot}$. This is the reason we analyze binaries with component masses between $1-5 M_{\odot}$. Since TD has little influence above $2M_{\odot}$ it is left to TH to help recognize the presence of horizons. We find that it is highly improbable to do so for binaries with component masses in the range $2-5 M_{\odot}$ in the current generation of detectors.

It has been shown that in GR the Love number vanishes for BHs, but not  for other compact objects like NSs~\cite{Binnington:2009bb, Landry:2014jka, Chia:2020yla}. However, recently it has been suggested that the Love number can be nonzero for  nonaxisymmetrically perturbed rotating BHs~\cite{LeTiec:2020bos}.  In the current work we  take that the tidal deformability of all BHs is zero. Thus, our results may need to be revisited depending on how this matter gets resolved. 

We begin by studying in Sec.~\ref{sec:TH} the TH terms that appear in the GW phase of a binary. There we identify two horizon parameters that are best measured for stellar-mass binary signals in ground-based detectors. There we also show how the spin-induced quadrupole moment and tidal deformability of the binary components influence the waveform. 
In Sec.~\ref{sec:bf} we develop the method for weighing the
evidence in data for the presence or absence of horizon, utilizing the aforementioned horizon parameters and phase terms in a Bayesian formalism.
In Secs.~\ref{sec:priors} and \ref{simulation} we implement this formalism on a large population of simulated binary signals in noisy data simulated with Advanced LIGO and Advanced Virgo noise. 
We conclude with a discussion on future prospects in Sec.~\ref{sec:conclusion}.

\section{Effect of TH on binary waveforms}
\label{sec:TH}

Consider a compact binary with component masses $m_i~~(i=1,2)$, 
total mass $m= m_1 + m_2$, and mass-ratio
$q=m_2/m_1$, with $m_2 \leq m_1$.
Let the dimensionless component spins be $\chi^{}_i$. Under the adiabatic approximation the orbital evolution of the binary can be quantified in the post-Newtonian formalism with reasonable accuracy, especially, when it is far from merger~\cite{Blanchet:2013haa}. The dynamics of the system is governed by energy and angular momentum loss from the orbit. Usually it has a contribution arising from taking the components as point particles (PP) and another one originating from their finite size. The latter can be decomposed into two parts, (i) tidal deformation of each component due to the gravitational field of the other and (ii) the amount of energy absorbed by individual components from the orbit, namely, {\em  tidal heating}.

The dynamics of the system and, therefore, the emitted GW depends on all of these contributions. Hence, the Fourier transformed GW waveform can be written as   
\begin{equation}
    \Tilde{h}(f) =  A(f) e^{i\left(\Psi_{\rm PP}+\Psi_{\rm TD}+\Psi_{\rm TH}\right)}\,,
\end{equation}
where $f$ is the instantaneous GW frequency and $A(f)$ is the frequency-dependent amplitude. The phase terms -- $\Psi_{\rm PP}, \Psi_{\rm TD},$ and $\Psi_{\rm TH}$ -- are the  phase contributions arising from the point-particle approximation, TD, and TH, respectively.

Since GW absorption is negligible for matter~\cite{Glampedakis:2013jya},
it is reasonable to exploit evidence of TH in binary waveforms to discern the existence of  horizons~\cite{Datta:2019euh, Maselli:2017cmm}. This expectation led us to introduce the {\em horizon} parameter $H$ for extreme mass-ratio inspirals (EMRIs) that LISA may observe~\cite{Datta:2019euh}. 
Till now horizon distinguishability employing TH has  been addressed primarily for LISA sources, such as EMRIs and supermassive BH binaries. However,
even in the case of supermassive BH binaries, it is the combined tidal heating of both binary components is what has been employed,
which ignores the possibility that not both components may have horizons (or lack them)~\cite{Maselli:2017cmm}. Such an approach is  reasonable for initial forays in this subject  but, in general, different values of $H$ need to be considered. 
Here we apply the formalism to binaries with similarly massive components primarily to target the LIGO-Virgo population of stellar-mass binaries. As it is a broad subject, we keep the studies with third-generation for the future.

For a near-equal-mass binary we define horizon parameters for each component, $(H_1, H_2)$, such that the value of $H_i$ is 1 (0) when the $i$th component has a horizon present (absent). In the case of circular orbits, the flux of energy at the horizon can be expressed as a PN expansion~\cite{Alvi:2001mx, Poisson:2018qqd, Poisson:2009di, Nagar:2011aa, Bernuzzi:2012ku,Chatziioannou:2016kem, Cardoso:2012zn}. Since TH signifies presence of horizon,
we multiply the  energy flux absorbed by each component with the corresponding $H_i$. In the case of partial absorption, one has $0<H_i<1$. Therefore, the absorbed flux is
\begin{equation}
\begin{aligned}
    -\frac{dE}{dt} = &{}\frac{32}{5}\nu^2 \frac{v^{15}}{4}\sum_{i=1}^{2} H_i\left(\frac{m^{}_i}{m}\right)^3 \left( 1 + 3\chi^2_i\right)\left\{-(\hat{L}_N.\hat{S}_i)\chi^{}_i \right.\\
    &\left. + 2 \left[ 1+\left( 1 - \chi^2_i\right)^{1/2}\right]\frac{m^{}_i}{m}v^3\right \}\,,
\end{aligned}
\end{equation}
where 
$\nu = {m_1 m_2}/{m^2}$ is the symmetric mass-ratio,  $v$ is the orbital velocity, and
$\hat{S}^{}_i$ and $\hat{L}_N$ are the unit vectors along the directions of the $i$th spin and the orbital angular momentum, respectively.

\subsection{ New waveform parameters characterizing TH}

\begin{figure}[htb!]
    \centering
    \includegraphics[width=\linewidth]{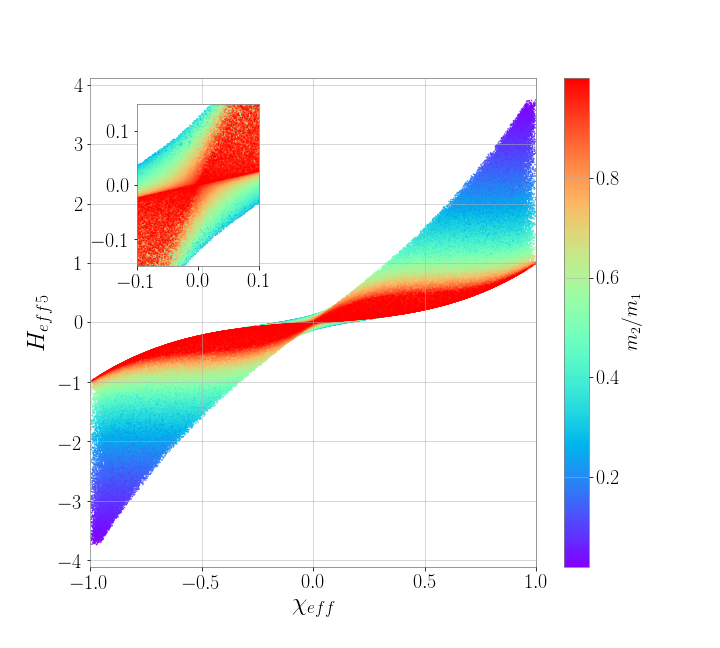}
    \caption{
    $H_{eff5}$ is plotted for a range of $\chi_{eff}$ values and for all possible values of $m_2/m$.}
    \label{Heff5_spin}
\end{figure}

\begin{figure}[htb!]
    \centering
    \includegraphics[width=\linewidth]{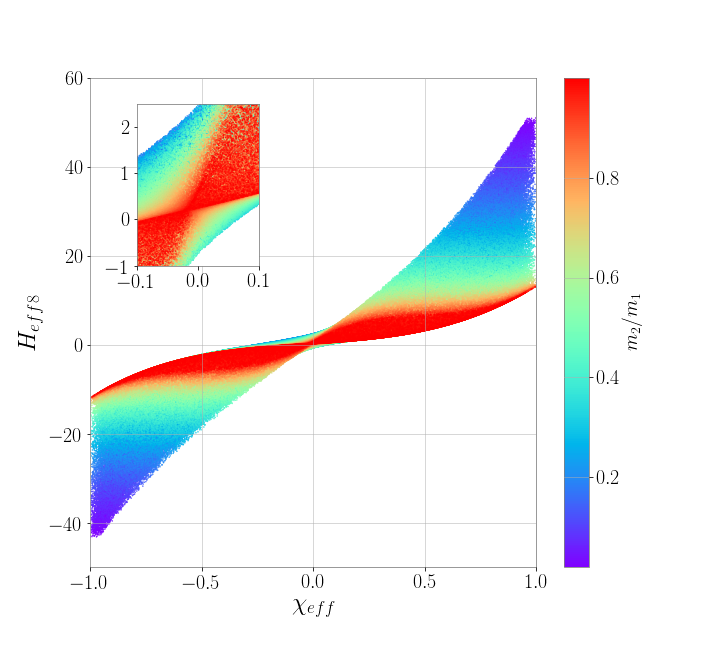}
    \caption{$H_{eff8}$ is plotted for a range of $\chi_{eff}$ values and for all possible values of $m_2/m$.}
    \label{Heff8_spin}
\end{figure}

The horizon parameters $H_{1,2}$ appear in the GW phase in terms that also include mass and spin factors. This makes them {\em degenerate} with those parameters, in that it is more practical to measure 
the following effective observables instead of $H_{1,2}$:
\begin{subequations}
\label{Eq.Hparams}
\begin{align}
H^{}_{eff5} \equiv &{} \sum_{i=1}^{2}H^{}_i \left(\frac{m^{}_i}{m}\right)^3 \left(\hat{L}.\hat{S}^{}_i\right)\chi^{} _i \left(3 \chi^{}_i{}^2+1\right)\,,\\
H^{}_{eff8} \equiv &{} ~4 \pi  H^{}_{eff5}+\sum^2_{i=1}H^{}_i \left(\frac{m_i}{m}\right)^4 \left(3 \chi^{}_i{}^2+1\right)\nonumber \\
                &\quad\quad\quad\quad\quad\quad\quad \times \left(\sqrt{1-\chi^{}_i{}^2}+1\right)\,.
\end{align}
\end{subequations}
These are analogous to the effective spin parameter $\chi^{}_{\rm eff}$ that was introduced~\cite{Damour:2001tu, Racine:2008qv,Ajith:2011ec} 
to characterize spinning compact binary waveforms: While the spins of the individual binary components are themselves difficult to measure (like $H_{1,2}$ here), their combined impact on the waveform phase, captured by $\chi^{}_{\rm eff}$, lends itself to more precise measurements. 
Dependencies of  $H_{eff5}$ and $H_{eff8}$ on component spins are shown in Fig.~\ref{Heff5_spin} and Fig.~\ref{Heff8_spin}, respectively.

If the system is a binary black hole (BBH), 
as long as any one of the component spins is finite both $H_{eff5}$ and $H_{eff8}$ will be nonzero.
By contrast, for the same spins a horizonless binary would have both $H_{eff5}$ and $H_{eff8}$ vanish. Therefore, it is easiest to discern between the presence and absence of horizons in BBHs that have at least one component with sufficiently large spin.

On the other hand, when both component spins of a BBH tend to zero, one has $H_{eff5} \to 0$ but  $H_{eff8}\neq 0$; see the inset in  Fig.~\ref{Heff8_spin}.
Therefore, in the low-spin limit $H_{eff8}$ emerges as a discriminator for the presence or absence of horizons. Here the measurement is helped for small mass-ratio ($q$), which ensures large $H_{eff8}$.

It is important to note that our choice of waveforms, based on the stationary-phase approximation (SPA)~\cite{Cutler:1994ys}, is for illustrative purpose, essentially as a proof of principle that the method proposed here is promising for identifying binary components with horizons from those without. For making such classification in real data, it will likely be important to use more accurate templates, such as those based on the EOB-NR formalism~\cite{Husa:2015iqa, Khan:2015jqa, Hannam:2013oca}. 
We will present those results in future. Having said that, our choice of SPA-based inspiral waveforms is a reasonable one for illustrating the power of this method for the systems studied here.

We deduce the GW phase involving TH by using Eq.~(2.7) of Ref.~\cite{Tichy:1999pv} (see \cite{Isoyama:2017tbp} for the details). We find  the phase shift due to the associated horizon absorption to be

\begin{equation}
\begin{aligned}
\label{eq:phase correction}
\Psi^{}_{\rm TH} = &{} \frac{3}{128\nu} \left(\frac{1}{v}\right)^5 \left[-  \frac{10 }{9 }v^5 H^{}_{eff5} \left(3 \log \left(v\right)+1\right) \right. \\
&-  \frac{5}{168} v^7 H^{}_{eff5} \left(952 \nu +995\right) \\ 
&\left.+   \frac{5}{9}v^8 \left(3 \log \left(v\right)-1\right)(-4 H^{}_{eff8}+ H_{{eff5}} \psi^{}_{\text{SO}} )\right]\,,
\end{aligned}
\end{equation}
where 
\begin{equation}
   \begin{split}
        \psi^{}_{\text{SO}} \equiv~ &\frac{1}{6} \big[\big(-56 \nu -73 \sqrt{1-4 \nu }+73\big) \big(\hat{L}.\hat{S}^{}_1\big) \chi _1 \\
        &+\big(-56 \nu + 73 \sqrt{1-4 \nu }+73 \big) \big(\hat{L}.\hat{S}^{}_2 \big) \chi _2\big]\,.
   \end{split}
\end{equation}
Note that $H_{eff5}$ and $H_{eff8}$ arise at different PN orders 
in the phase.


\subsection{New waveform parameters characterizing quadrupole moment}

The two bodies in a coalescing compact binary can have spin. In case of a nonzero spin a body would develop a spin-induced quadrupole moment.
The leading order contribution arises due to the mass quadrupole moment $(M_{2(i)})$ of both bodies, $(i=1,2)$, at 2 PN order. If the bodies are BHs, then $M_{2(i)} = -\chi^2_{(i)} m_{(i)}^3$. If they are NSs or ECOs that moment may be modified as $M_{2(i)} = -\kappa_{(i)} \chi_{(i)}^2 m_{(i)}^3$. 
Measuring the quadrupole moment $\kappa$ from observations can be used to probe the nature of the compact objects~\cite{Krishnendu:2017shb, Krishnendu:2018nqa, Datta:2019euh, Krishnendu:2019ebd}. Since in a binary the quadrupole moments $\kappa^{}_i$ of both the bodies  contribute at the similar order, they are degenerate. Usually, a combination of $\kappa^{}_1$ and $\kappa^{}_2$ are used for the measurement~\cite{Krishnendu:2017shb}. Here we define a new effective parameter $\tilde{Q}$ as follows:
\begin{equation}\label{Q_tilde_def}
 \tilde{Q} m^4= -\kappa^{}_+S_{\ell}^2 + S_{\ell}\Sigma_{\ell}(\kappa^{}_- - \delta\kappa^{}_+) + \frac{\Sigma_{\ell}^2}{2} (\delta \kappa^{}_- -\kappa^{}_+ +2\nu\kappa^{}_+)\,,
\end{equation}
where $\kappa^{}_+  = \kappa^{}_1 + \kappa^{}_2$, $\kappa^{}_- = \kappa^{}_1 - \kappa^{}_2$, $\delta = \frac{m^{}_1-m^{}_2}{m}$, $S_{\ell} = \sum_i \hat{L}.\hat{S}_i \chi^{}_i m_i^2$, $\Sigma_{\ell} = m (\hat{L}.\hat{S}_2 \chi^{}_2 m^{}_2 - \hat{L}.\hat{S}_1 \chi^{}_1 m^{}_1)$. Then the phase can be expressed as~\cite{Bohe:2015ana}
\begin{equation}\label{Q_tilde_phase}
\psi_{QM} = -\frac{25}{32}\frac{\tilde{Q}}{\nu }\frac{1}{v} \,.
\end{equation}
Note that once the phase has been expressed in terms of $\tilde{Q}$, it is not necessary for $\tilde{Q}$ to be limited
to Eq.~(\ref{Q_tilde_def}): It is straightforward to incorporate other models of $M_2$ into it, such as for boson stars~\cite{Ryan:1996nk, Cardoso:2019rvt}. We will use Eq.~(\ref{Q_tilde_phase}) in our modeling of phase due to nonzero quadrupole moment.
In Refs.~\cite{Krishnendu:2019tjp, Narikawa:2021pak}, observational constraint on spin induced quadrupole moment has been found. Individual measurement suffers from broad posterior distribution, pointing towards low measurability with current detectors.
Our results below are consistent with these observations.

\section{Bayes factors for horizon discrimination}
\label{sec:bf}

In the presence of a GW signal strain $h(t, \pmb{\theta})$, characterized by parameters $\pmb{\theta}$, 
the detector strain time-series can be modeled as $d(t) = n(t) + h(t, \pmb{\theta})$, where $n(t)$ denotes the detector's noise. In the presence of a GW signal $h(t,\pmb{\theta})$, described by a model $\mathcal{H}$, the likelihood of the data  is~\cite{Cutler:1994ys}:
\begin{equation}
\label{eq:likelihood}
P\left(d|\pmb{\theta}, \mathcal{H} \right) \propto \exp{ \left[ -\frac{1}{2} \langle d - h(\pmb{\theta})| d - h(\pmb{\theta})  \rangle \right]}\,,
\end{equation}
under the assumption of Gaussian and stationary-detector noise.
The angular bracket in Eq.~(\ref{eq:likelihood}) defines a noise-weighted inner product between two real time-series $a(t),\, b(t)$, and is given as 
\begin{equation}
    \langle a, b \rangle =  4\Re \int_{f^{}_{low}}^{f^{}_{high}} df \frac{\tilde{a}^*(f)\tilde{b}(f)}{S_n(f)},
\end{equation}
where $S_n(f)$ is the one-sided power spectral density (PSD) of the detector noise, and $f^{}_{low}$ and $f^{}_{high}$ are the low-frequency cutoff and high-frequency cutoff, respectively~\cite{Cutler:1994ys, Allen:2005fk}. Using the inner product, one can also define the signal-to-noise ratio (SNR) $\rho$ for the template $h(t, \pmb{\theta})$ as
\begin{equation}
    \rho = \frac{\langle d|h \rangle}{\sigma},
\end{equation}
where $\sigma = \sqrt{\langle h|h\rangle}$ is the template normalization.

We will assume that noncolocated detectors on the globe have uncorrelated noise; hence, the combined likelihood is given as~\cite{Veitch:2014wba},
\begin{equation}
    P\left(\mathbf{d}|\pmb{\theta}, \mathcal{H} \right) =  \prod_i^N  P\left(d_i|\pmb{\theta}, \mathcal{H} \right)\,,
\end{equation}
where $\mathbf{d}\in\{d_1, d_2,\cdots,d_N\}$ represents combined data from all $N$ detectors. Using the coherent network likelihood function, posterior probability density can be written as 
\begin{equation}
    P\left(\pmb{\theta}|\mathbf{d},\mathcal{H} \right) = \frac{P\left(\mathbf{d}|\pmb{\theta}, \mathcal{H} \right) P\left(\pmb{\theta}|\mathcal{H} \right)}{ P\left( \mathbf{d}|\mathcal{H}\right)}\,,
\end{equation} 
where $P\left(\pmb{\theta}|\mathcal{H} \right)$ is the prior probability density function or prior of the parameters
$\pmb{\theta}$. In the denominator, $P\left( \mathbf{d}|\mathcal{H}\right)$ is the marginalized posterior probability density over all parameters $\pmb{\theta}$, and is also known as the evidence for the model $\mathcal{H}$. 
The evidence $P\left(d|\mathcal{H} \right)$ serves as a normalization constant of the posterior probability for $\mathcal{H}$. 
The evidence computed for two competing models or hypotheses can be used to determine which one is favored by the data.  
In this work, we compute Bayes factors for simulated signals to compare two hypotheses, namely,
\begin{enumerate}
    \item The horizon hypothesis $\mathcal{H}_\textrm{BBH}$: Signal carries imprints of horizon  absorption and  spin-induced quadrupole moment,
    \item  The no-horizon hypothesis $\mathcal{H}_{\textrm{BNS}}$: Signal has no imprint of  horizon absorption, but has TD and  spin-induced quadrupole moment.
\end{enumerate}

In Bayesian model selection, we compute  the Bayes factor,
\begin{equation}
\label{eq:BF}
\textrm{BF} = \frac{P\left( \mathbf{d}|\mathcal{H}_\textrm{BBH}\right) }{P\left( \mathbf{d}|\mathcal{H}_{\textrm{BNS}}\right)}\,.
\end{equation}
If the Bayes factor is greater than some preset threshold, {\it i.e.}, $\textrm{BF}> \textrm{BF}_{\textrm{Th}}$ then the hypothesis $\mathcal{H}_\textrm{BBH}$ is preferred  over the other hypothesis $\mathcal{H}_\textrm{BNS}$ in the data.
Moreover, we use the Dynesty sampler~\cite{2020MNRAS.tmp..280S}, as implemented in the Bilby package~\cite{Ashton:2018jfp, Romero-Shaw:2020owr}, to compute the posterior probability densities for our simulated signals.  
We use a likelihood function marginalized over time $t_c$ and phase $\phi_c$ at coalescences of binaries~\cite{time-phase-marginalization,Thrane:2018qnx} and distance $d_L$~\cite{Singer:2015ema,Singer:2016eax},  thus removing the need for
sampling those parameters without affecting the posterior probability densities in the parameters of interest. The posterior probability densities for these parameters can be reconstructed analytically from the full set of posterior samples~\cite{Thrane:2018qnx}.

The posteriors of 
some of the parameters for the $\mathcal{H}_\textrm{BBH}$ hypothesis 
are shown in Fig.~\ref{fig:horizon_parameters}. To compute them, we considered the signal integration in a frequency range such that it ends at $f_{\rm ISCO}$, while the duration of the signal is 16s. $f_{\rm ISCO}$ is the instantaneous GW frequency at the ISCO of the binary~\cite{Kidder:1992zz,Blanchet:2001id}. In practice, it may be possible to begin the signal integration at a frequency as low as 10Hz, which is what aLIGO design targets. Similarly, when waveform modeling is available to accurately incorporate  TH beyond the ISCO, the upper frequency cutoff will also be raised. Both these changes will improve parameter estimation as well as Bayes-factor based model discrimination.

Before setting up signals simulation for BF computations, it is worthwhile to examine through computationally inexpensive, even if approximate, means how precisely the horizon parameters would be measurable in mass-gap binaries. Such a computation is afforded by the Fisher information matrix (FIM), as defined below.
We estimate how large the noise-limited errors are of the horizon parameters ${\boldsymbol{\vartheta}}$, by modeling the measured values after the maximum likelihood estimators {MLEs}~\cite{Helstrom:1994}.
Owing to noise, the {MLE} will fluctuate about the respective true values, i.e., $\hat{{\boldsymbol{\vartheta}}} = {\boldsymbol{\vartheta}} + \delta{\boldsymbol{\vartheta}}$, where $\delta{\boldsymbol{\vartheta}}$ is the random error.
The extent of these fluctuations is estimated by the elements of the variance-covariance matrix, $\gamma^{ab}=\overline{\delta{\boldsymbol{\vartheta}}^a \delta{\boldsymbol{\vartheta}}^b}$~\cite{Helstrom:1994},
which is bounded by the signal via the Cramer-Rao inequality, namely,
\begin{equation}
  \norm{\mat{\gamma}} \geq \norm{\mat{\Gamma}}^{-1}\;,
\end{equation}
where $\mat{\Gamma}$ is the FIM:
\begin{align}
  \label{eq:Fisher}
  \Gamma_{ab}
   & = \left\langle{\partial_a \tilde{h}({\boldsymbol{\vartheta}}), \partial_b \tilde{h}({\boldsymbol{\vartheta}})}
   \right\rangle \,, \nonumber                                 \\
   & \equiv  4 \Re \int_{f_{low}}^{f_{high}} d{f}\,\frac{\partial_a \tilde{h}^{*}(f; {\boldsymbol{\vartheta}})\;\partial_b \tilde{h}(f; {\boldsymbol{\vartheta}})}{S_{n}(f)}\,.
\end{align}
Above, $\partial_a$ is the partial derivative with respect to the parameter $\vartheta^a$. 
Therefore, $\Delta \vartheta^a \equiv (~{\overline{\delta \vartheta^a\,\delta \vartheta^a}}~)^{-1/2} = \Gamma_{aa}^{-1/2}$ gives the lower bound on the root-mean-square error in the estimate of $\vartheta^a$.
The two are equal in the limit of large {SNR}~\cite{Helstrom:1994}.
The error estimates listed here are the $\Delta\vartheta^a$ obtained from the {FIM}.

When one computes 
$\Gamma_{ab}$ for the binary parameters one typically finds that its offdiagonal terms are nonzero, which implies that there are covariances among the parameter errors. It is, however, possible to mitigate those covariances for a different set of parameters. In the two-dimensional parameter subspace of $(H_1,H_2)$, we find that $(H_{eff5},H_{eff8})$ are such parameters.

One can also use FIM to deduce errors $(\Delta H_{eff5},\Delta H_{eff8})$ in the new horizon parameters for our binaries of interest. This is how we estimate that for a mass-gap binary at a distance of 10Mpc to a few tens of Mpc, it is possible to measure
$( H_{eff5},H_{eff8})$ to a few tens of percent in a three detector LIGO-Virgo network with the aforementioned noise PSD. A similar FIM calculation for the third generation detector Einstein Telescope shows that the same  measurement precision is achievable even when the same mass-gap BBH is pushed out to a few 100 Mpc. 
As mentioned above, in spite of the weak effect of TH in mass-gap binaries in current detectors,  for the completeness of the waveforms used in our simulations we continue to retain the TH terms in their phases. The impact of those terms for third generation detectors and binaries not limited to the mass-gap will be studied elsewhere.

\section{Priors}
\label{sec:priors}

The distributions and  ranges of parameter priors of the simulated binary waveforms used in our Bayesian model selection studies are listed in  Table~\ref{tab:priors}. The possible values of $H_{eff5}$ and $H_{eff8}$ are shown in Fig.~\ref{Heff5_spin} and Fig.~\ref{Heff8_spin}, respectively.  

\begin{figure}
    \centering
    \includegraphics[width=\linewidth]{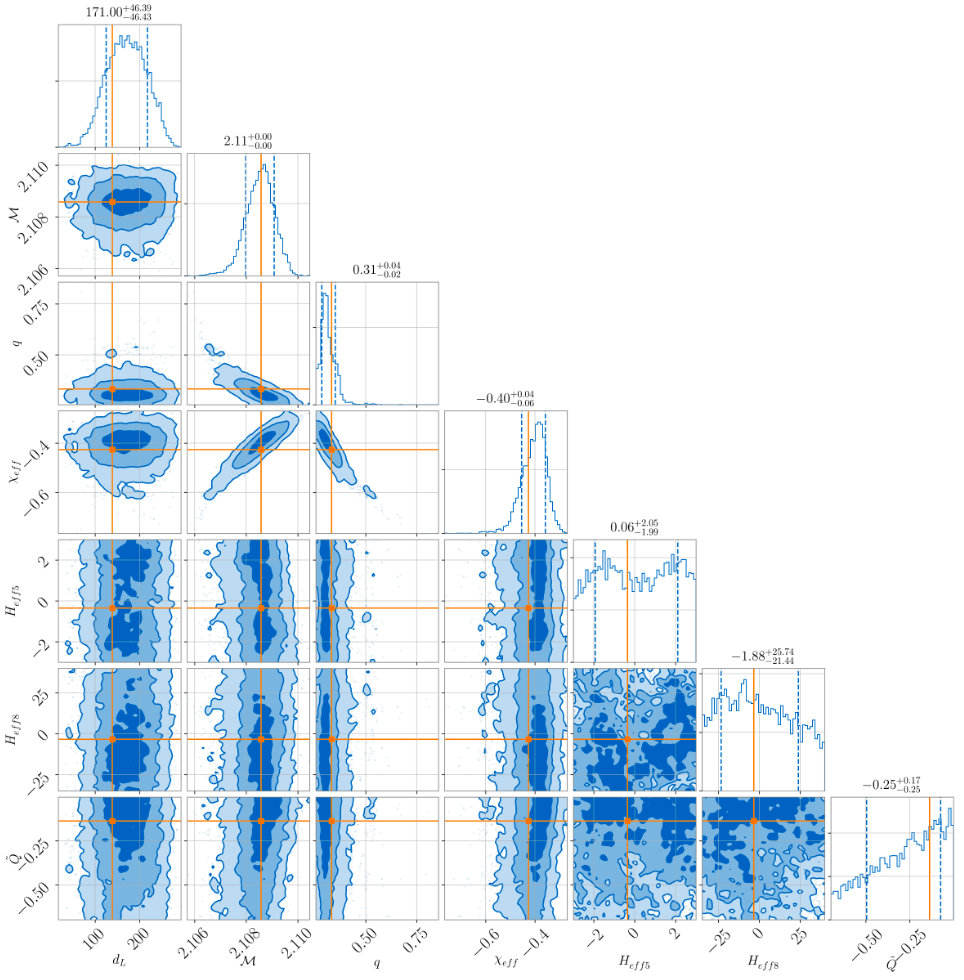}
    \caption{This parameter estimation corner plot shows the posterior probabilities of the parameters of a BBH injection. The injected values are indicated by the orange lines. The posteriors   show that the measurement of the new parameters introduced here,  $H_{eff5}$ and $H_{eff8}$, will add very little bias to the estimation of standard intrinsic parameters of compact binaries. Those parameters are estimated as precisely as expected from past studies.
    }
    \label{fig:horizon_parameters}
\end{figure}

In Fig.~\ref{fig:horizon_parameters}, we show the posterior probability distributions of various parameters of a BBH injected signal obtained from 
a Bayesian analysis. The luminosity distance $(d_L)$, chirp mass $(\mathcal{M})$, mass ratio $(q)$, and effective spin $(\chi_{eff})$ are well measured. The estimation recovers the injected values. Comparatively $H_{eff5}$ and $H_{eff8}$ are poorly measured. Although we recover the injected values, and the posterior is certainly different from the flat prior, the error is large. This is expected as TH is a higher-order effect.

\begin{table*}
\caption{Choice of priors in our Bayesian analysis of simulated signal injections.  \label{tab:priors}}
\begin{ruledtabular}
\begin{tabular}{c c c c c}
{Parameter} & {Distribution} & {Range} & {Boundary condition} & Units \\ \hline
Chirp mass ($\mathcal{M}$) & Uniform & [1.0, 4.5]  & -- & $\textrm{M}_\odot$\\
Mass ratio ($q$) & Uniform & [0.5, 1.0] & --& --\\
Spin of primary object $(\chi_1)$ & Uniform & [0.00, 0.99] & Reflective & --\\
Spin of secondary object $(\chi_2)$ & Uniform & [0.00, 0.99] & Reflective & --\\
Tidal deformability of primary object $(\Lambda_1)$ &Uniform & [0.0, 500]  & -- & --\\
Tidal deformability of secondary object $(\Lambda_2)$ &Uniform& [0.0, 800]  & -- & --\\
2.5 PN horizon parameter ($H_{eff5}$) & Uniform & [-4.0, 4.00] & -- & --\\
4 PN horizon parameter ($H_{eff8}$) & Uniform & \!\,[-45,  45.0] & -- & -- \\ 
Luminosity distance ($d_L$) & Uniform & \!\!\!\,[10.0, \;500] & -- & Mpc\\
Right ascension (RA) & Uniform & \!\![0.0, $2\pi$] & Periodic & radian \\
Declination (DEC) & Cosinusoidal & \![$-\pi/2$, $+\pi/2$] & --& radian \\
Phase at reference frequency $(\phi^{}_0)$ & Uniform & \,[0.0, $2\pi$] & Periodic & radian\\
Polarization angle $(\psi)$ & Uniform & \;[0, $\pi$] & Periodic & radian\\
Line-of-sight angle $(\theta_{JN})$ & Sinusoidal & \;[0, $\pi$] & -- & radian \\
Effective parameter for spin-induced quadrupole moment $\tilde{Q}$ & Uniform & [$-2$, 0.0] & --& --
\end{tabular}
\end{ruledtabular}
\end{table*}

\section{Simulation and Results}
\label{simulation}

\subsection{Properties of the simulated sources}

We now quantify how successfully one can discriminate between a BBH signal from a BNS one in noisy data. For this {\em signal model selection test} we simulated a population of 1250 binaries, which are distributed uniformly in comoving volume between 50Mpc to 250Mpc. Component masses were taken to be $m^{}_{1,2} \in~[1-5] M_{\odot}$ and spins chosen to be aligned or antialigned with the orbital angular momentum,
and with dimensionless magnitude $\chi^{}_{1,2} \in~[0, 0.9]$. 
For model selection we  constructed two families of waveforms -- both for signals (for adding in simulated noisy data) and templates (for matched-filtering that data) -- namely: (a) TaylorF2 (TF2), modified with   TD contribution (TidalTF2) for representing horizonless components with nonzero   TD. Here, the GW phase is devoid of any contribution from $H_{eff5}$ or $H_{eff8}$; (b) HeatedTaylorF2 (HTF2), which is TF2 but with additional phase terms arising from  TH, as described in Eq.~(\ref{eq:phase correction}). We have included the effect of the spin-induced quadrupole moment appropriately in both cases via the phase term in Eq.~(\ref{Q_tilde_phase}). We used the Akmal, Pandharipande, and Ravenhall (APR) equation of state (EOS)~\cite{Akmal:1998cf} for this purpose \cite{Pappas:2012qg} to model the new effective parameter $\tilde{Q}$ introduced in Eq.~(\ref{Q_tilde_phase}). The injected values of $\tilde{Q}$ have the range $[-1.60,0]$.
From Ref. \cite{Pappas:2012qg} we constructed the values of $\kappa^{}_i$ of the $i$th body of mass $m_i$ and spin $\chi^{}_i$. From these values we find the corresponding value of $\tilde{Q}$ using Eq. (\ref{Q_tilde_def}), which is used for injection.

Using the aforementioned waveform models we performed simulated signal injection studies in simulated colored-Gaussian data of two LIGO detectors (Hanford and Livingston) with aLIGO zero-detuned high-power (ZDHP) noise power-spectral density~\cite{design-sensitivity}. To keep computational costs manageable we limited all our signals (and the filtering and parameter estimation) to only 16sec, and till the innermost stable circular orbit (ISCO). In one study, the sources are taken to be CBCs, with BHs as components. Hence, the injected waveform used is HTF2. We then performed a Bayesian analysis to measure the  parameters of these sources with both TidalTF2 and HTF2 templates and compared the natural-log of their Bayes factor $\ln{\mathrm{BF}}$, utilizing the definition in Eq.~\ref{eq:BF}, for the same ``horizon" injections to test if such an analysis has the power to identify the true signal model.

\begin{figure}
    \centering
    \includegraphics[width=\linewidth]{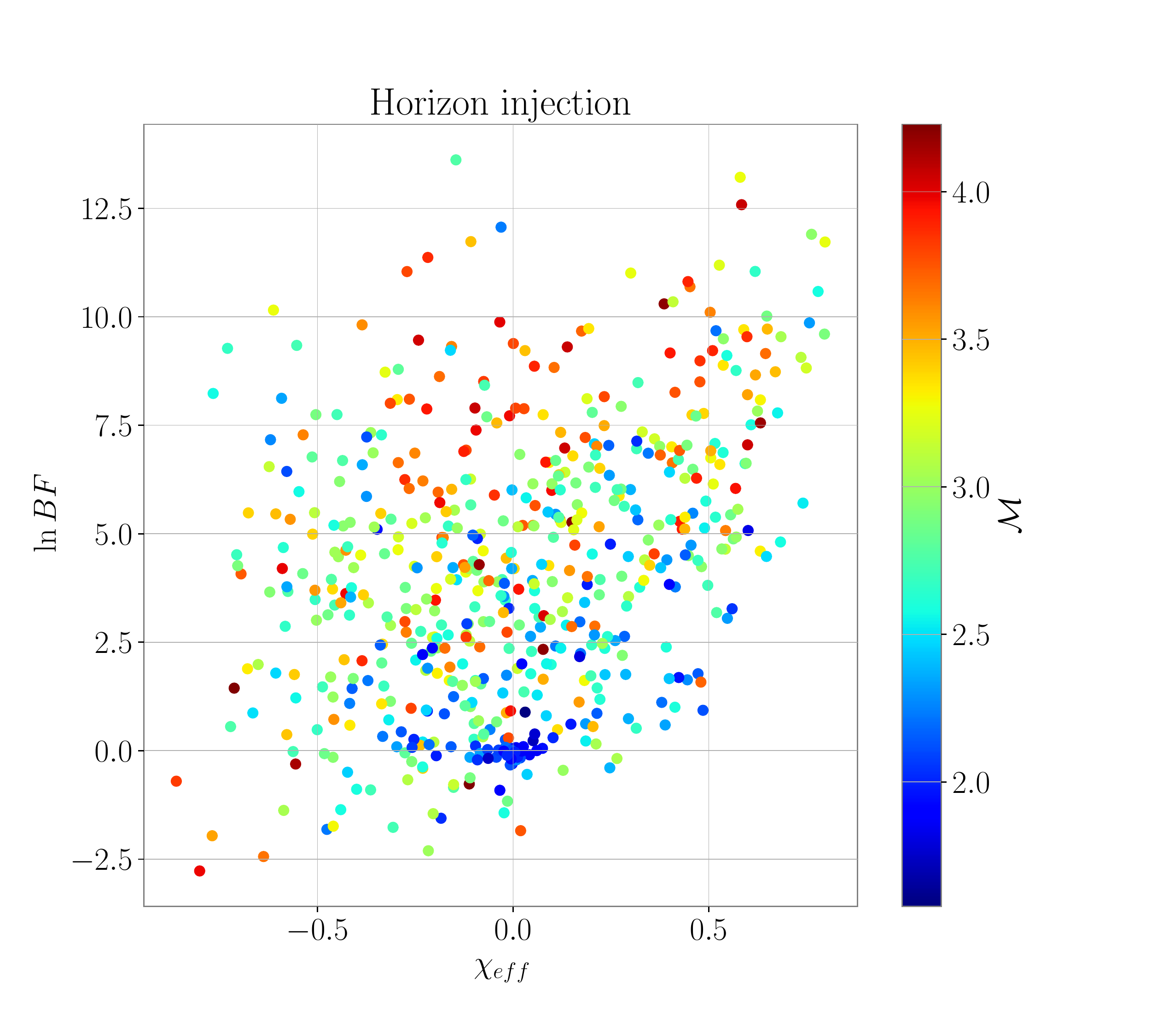}
    \caption{The logarithms of Bayes factors are plotted {\em vs} $\chi_{eff}$  above, in support of HTF2, when the injection is also HTF2.  The color bar represents the chirp masses of injections. \label{fig:BF_wrt_m_s} 
}

\end{figure}

In Fig. \ref{fig:BF_wrt_m_s} we plot the $\ln{\rm BF}$ with respect to $\chi_{eff}$ in  the $x$-axis and $\mathcal{M}$ in color. Each point in this figure represents an HTF2 injection. The fact that for a large majority of them the values of  $\ln{\rm BF}$ are positive, suggests that for this injection set model selection strongly favors horizon injections. 
In an ideal case, all of the points should be above the $\ln {\rm BF}=0$ line. Deviation from this expectation for a minority of the injections is due to their low SNRs or similarity of their signals with the TidalTF2 waveforms for the same parameters (as will be explored in more detail below). With longer duration waveforms and higher SNRs this result should get somewhat better, for lower masses.
The origin of the high values is likely due to a combination of TD, quadrupole moment, which will be discussed below.

\subsection{Assessing the statistical significance of the horizon discriminator}
\label{sec:foreandback}

In the preceding section, we found that barring a small subset the model selection returns positive $\ln {\rm BF}$ values for the injected sources, 
which is tantamount to saying that the observations favor the true signal model, namely, HTF2 here.
However, for a small subset, with negative ln~BF values, the wrong signal model (TidalTF2) is preferred. This raises the possibility that the opposite can also happen, i.e., some TidalTF2 injections, 
searched with both types of templates, may return  $\ln {\rm BF}$ values favoring the HTF2 signal model. As with any statistical analysis, it also becomes important to interpret quantitatively the probability with which the nature of those sources will be identified correctly. BBH injection studies enable one to do precisely that.
However, it is also important to assess the probability with which the nature of that source will be misidentified. For example, if the value of $\ln{\mathrm{BF}}$ turns out to be 5 for a BBH signal, it is important to interpret that value in terms of how probable it is to be identified correctly  as of BBH origin (the true hypothesis, $\mathcal{H}_{\rm BBH}$) and incorrectly from a BNS (the wrong hypothesis, $\mathcal{H}_{\rm BNS}$), in noisy data.
As we discuss next, the former probability can be assessed from the above study of BBH injections, whose $\ln{\mathrm{BF}}$ values form the {\em foreground} distribution 
$p\left(\ln \mathrm{BF}\mid \mathcal{H}_\textrm{BBH}  \right)$, which is the probability distribution of the ln~BF values given that the $\mathcal{H}_\textrm{BBH}$ hypothesis is true, i.e., the (injected) signals belong to the HTF2 model.
On the other hand, to assess 
how probable it is for the TidalTF2  signals to be misidentified as HTF2, we also study injections of horizonless signals generated using the TidalTF2 waveform model; these $\ln{\mathrm{BF}}$ values form the {\em background} distribution $p\left(\ln \mathrm{BF}\mid \mathcal{H}_\textrm{BNS}  \right)$ when the hypothesis being tested for a detected event is that it is from a BBH.

\begin{figure}
    \centering
    \includegraphics[width=\linewidth]{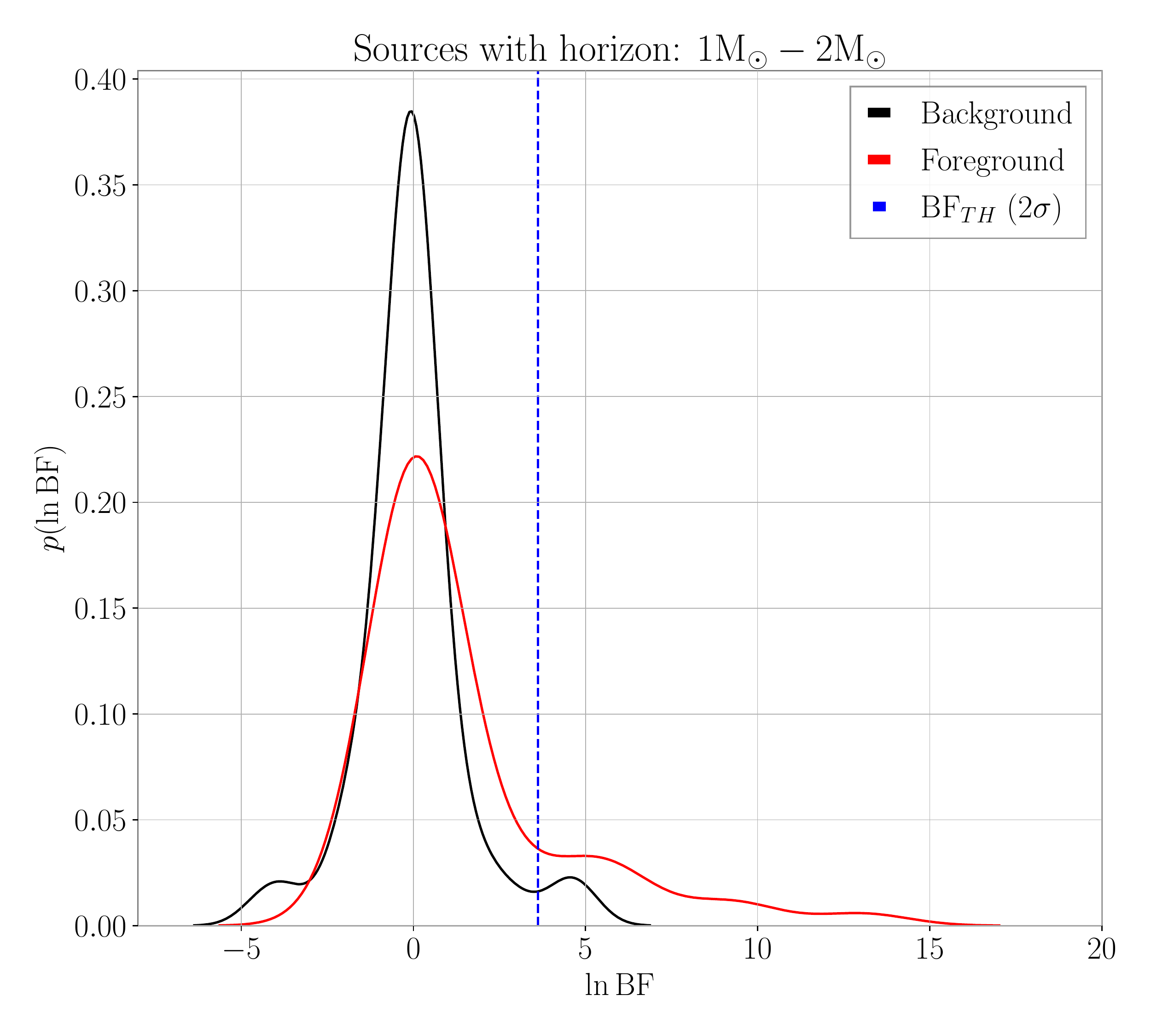}
    \caption{ \label{fig:background_and_threshold} The red curve represents the foreground distribution $p\left(\ln \mathrm{BF}\mid \mathcal{H}_\textrm{BBH}  \right)$ of $\ln{\mathrm{BF}}$ obtained from simulated BBH signal injections, with $m_i \in [1,2]M_\odot$, in simulated Gaussian noise of Advanced LIGO. The blue vertical line denotes threshold log Bayes factors $\ln{\mathrm{BF}}_{\mathrm{threshold}}$ corresponding to  $2\sigma$ significance. The black curve depicts the background distribution $p\left(\ln \mathrm{BF}\mid \mathcal{H}_\textrm{BNS} \right)$ obtained from BNS injections. 
}
\end{figure}

To obtain the foreground distribution corresponding to HTF2 signals, we compute the BF values for HTF2 injections using Eq.~(\ref{eq:BF}). We plot the distribution of the ln~BF for these values in red in Fig.~\ref{fig:background_and_threshold}.
To construct the background distribution for the same signals, we compute the BF values -- but now for {\em TidalTF2} injections -- using Eq.~(\ref{eq:BF}).
The distribution of the ln~BF for these values is plotted in black in the same figure.
The samples of $\ln \mathrm{BF}$ from the foreground and background distributions are used to estimate the efficiency with which BBHs can be identified in GW events and assign a statistical significance to each identification.

In Fig.~\ref{fig:background_and_threshold}, we show the estimated foreground and background  distributions of the $\ln {\rm BF}$ for a subpopulation of the injected binaries discussed 
in 
Fig.~\ref{fig:BF_wrt_m_s} and the preceding subsection. This subpopulation includes only those binaries that have component masses in the range $m^{}_{1,2}\in [1 - 2] \mathrm{M}_\odot$. As noted earlier, the foreground distribution is constructed by injecting HTF2 and calculating the Bayes factor in favor of HTF2.  The background distribution is estimated by calculating the Bayes' factor in support of HTF2 for the TidalTF2 injections.
We use the background distribution $p\left(\ln \mathrm{BF}\mid \mathcal{H}_\textrm{BNS}  \right)$ to compute the 
{\it false-detection-probability} (FDP) of a BBH claim, given that a BNS signal is actually present in the data.\footnote{The FDP is the frequentist {\it p}-value of hypothesis testing~\cite{Pearson1992,Neyman1992}.} The FDP is computed from   $p\left(\ln \mathrm{BF}\mid \mathcal{H}_\textrm{BNS}  \right)$ for a measured $\ln \mathrm{BF}$-value $\ln \mathrm{BF}_{\mathrm{measured}}$  as follows: 
\begin{equation}
    \mathrm{FDP} =  1 - \int_{-\infty}^{\ln \mathrm{BF}_{\mathrm{measured}}} p\left(\ln \mathrm{BF}\mid \mathcal{H}_\textrm{BNS}  \right) d\ln \mathrm{BF}.
\end{equation}
If the FDP is sufficiently low, then it is less likely that the event is consistent with the $\mathcal{H}_{\mathrm{BNS}}$ hypothesis. Often the FDP values are converted to equivalent significance levels, {\it e.g.}, $\mathrm{n}\,\sigma$ deviation of a Gaussian random process. From the background distribution of $\ln \mathrm{BF}$, we can compute the threshold Bayes' factor $\ln \mathrm{BF}_{\mathrm{threshold}}$ corresponding to a certain statistical significance.

In our analysis, each GW signal  is injected in a 16-second-long simulated colored-Gaussian data of the two  Advanced LIGO detectors.
In 
Fig.~\ref{fig:background_and_threshold}, the foreground distribution  $p\left(\ln \mathrm{BF}\mid \mathcal{H}_\textrm{BBH}  \right)$ and the background distribution  $p\left(\ln \mathrm{BF}\mid \mathcal{H}_\textrm{BNS}  \right)$ are shown in red and black colors, respectively. The blue vertical line denotes the threshold value $\ln{\mathrm{BF}}_{\mathrm{threshold}}$ corresponding to  $2\sigma$ significance. 
Above that $2\sigma$ threshold, the areas under the foreground and the background curves are $\sim 0.17$ and $\sim 0.04$, respectively. Therefore, around $17\%$ of 
BBH signals will have ln~BF greater than that threshold, and will be correctly identified as BBHs with a significance at the $2\sigma$ level.
However, there is a 4$\%$ chance that
a signal from a BNS will be mischaracterized as a BBH at the same level of significance.
We notice that there is a significant overlap between the foreground and background distributions, which is expected to decrease somewhat with longer waveforms and more sensitive detectors.

\begin{figure}
    \centering
    \includegraphics[width=\linewidth]{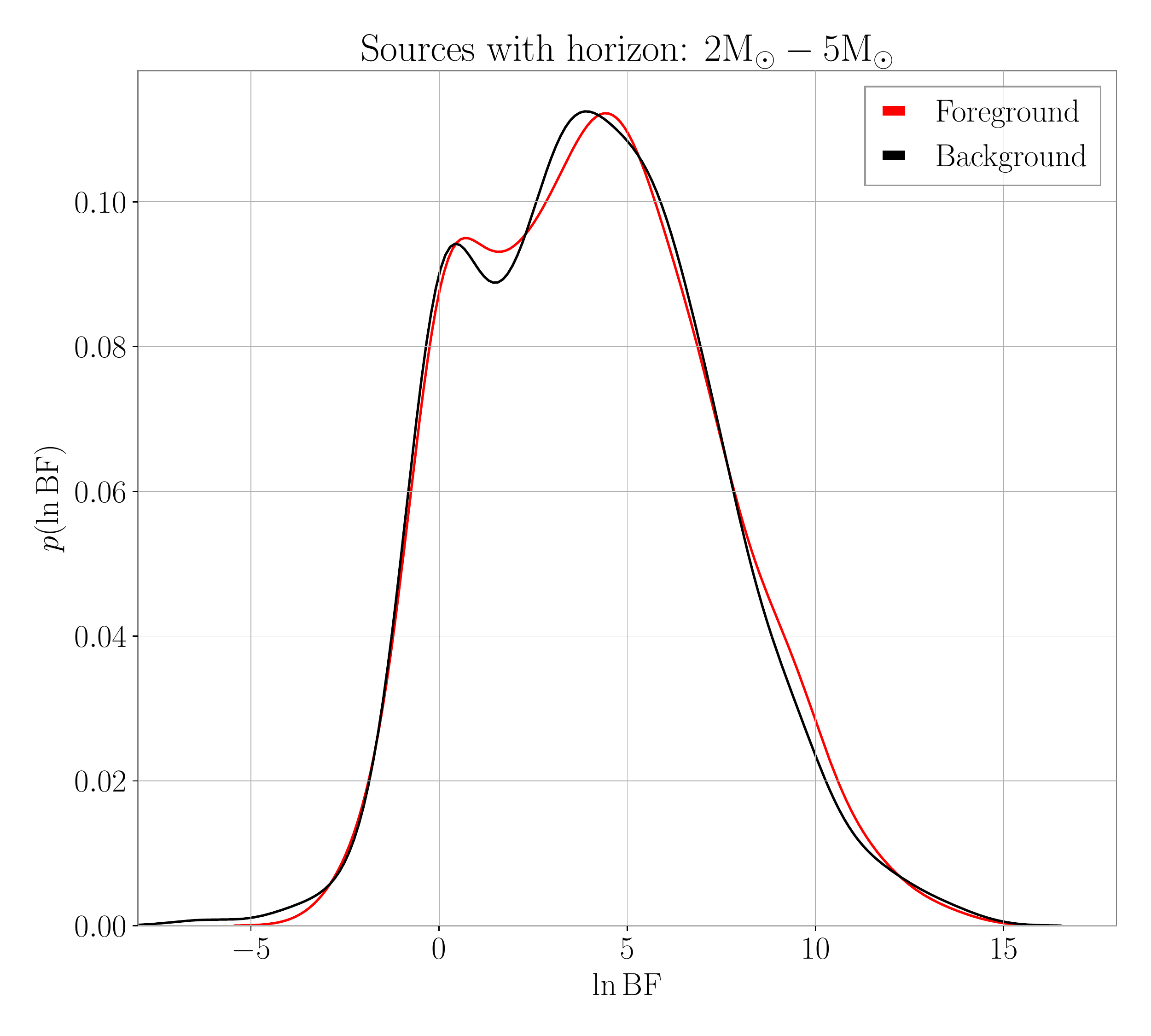}
    \caption{ \label{fig:m_grtr_2} Same as Fig.~\ref{fig:background_and_threshold}, but for BBH injections with component masses $m_i \in [2,5]M_\odot$. The red curve represents the foreground distribution $p\left(\ln \mathrm{BF}\mid \mathcal{H}_\textrm{BBH}  \right)$ of $\ln{\mathrm{BF}}$ obtained from BBH injection runs in the simulated Gaussian noise of Advanced LIGO. The black curve depicts the background distribution $p\left(\ln \mathrm{BF}\mid \mathcal{H}_\textrm{BNS}  \right)$ of $\ln{\mathrm{BF}}$ obtained from BNS injections.
}

\end{figure}

In Fig. \ref{fig:m_grtr_2} we perform the same exercise for heavier component masses, namely, $m_i>2 M_{\odot}$. In this mass range interestingly we find that irrespective of the type of the source injected, model selection prefers  HTF2. 
This is because for those masses neither the TD (and $\tilde{Q}$) phase terms in TidalTF2 nor the TH ones in HTF2 are large enough to induce  phase difference between the TidalTF2 and the HTF2 waveforms that is significant enough to tell them apart. In fact, both waveforms are very similar to the point-particle waveform there. 
While with increasing total mass TH would eventually become large, nevertheless in the mass-gap it is weak enough to tell such binaries apart at realistic distances in the current generation of detectors.

\begin{figure}
    \centering
    \includegraphics[width=\linewidth]{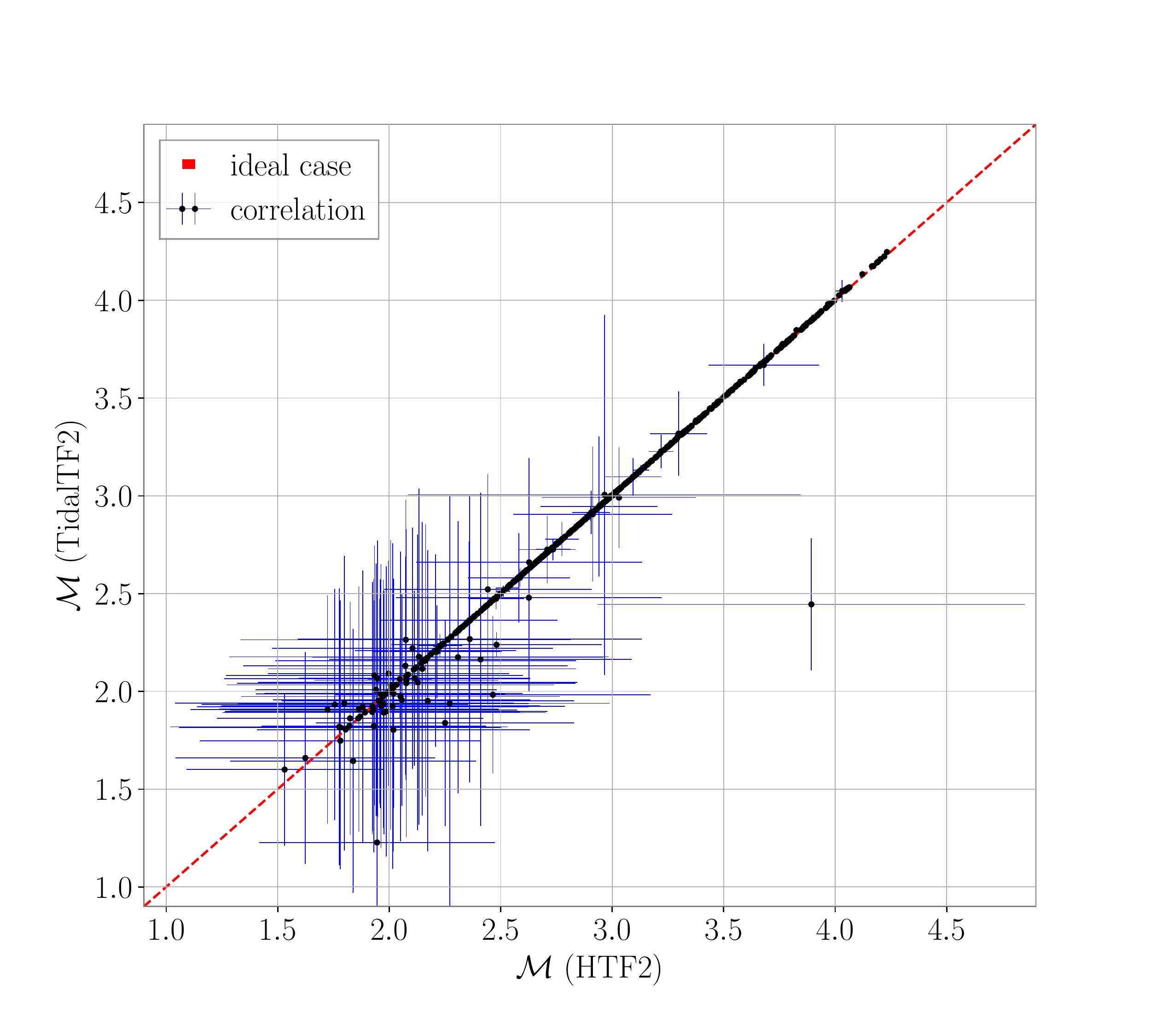}
    \caption{
   For TidalTF2
   signal injections, we do parameter estimation with both TidalTF2 and HTF2 waveforms and plot the measured values of the chirp mass $\mathcal{M}$ here. The range of the binary component masses range is $m_{1,2}\in [1,5]M_\odot$. In the absence of systematics the measured values should be highly correlated between TidalTF2 and HTF2 measurements, and should fall along the diagonal line. In the presence of noise, however, we expect some scatter around that line -- as evidenced here.  As expected, the measured $\mathcal{M}$ values shown above are nicely clustered around the diagonal. This shows that any biases, if present, are much less significant than the statistical errors.
    \label{fig:M_chirp_corelation} 
}
\end{figure}

We further test the conclusion above with a study of any systematics that may be induced in the estimation of parameters when the wrong waveform is used to search for a signal. For this purpose, we focus 
on the most precisely measured binary parameter, namely, the chirp mass ${\cal M}$. In Fig.~\ref{fig:M_chirp_corelation} we investigate these systematics. When the injection is TidalTF2 but parameter estimation employs  TidalTF2, on the one hand, and HTF2, on the other hand. We plot the values of $\mathcal{M}$ so measured in Fig.~\ref{fig:M_chirp_corelation}. If there were no systematics, then the measured values should be highly correlated between TidalTF2 and HTF2 measurements. In an ideal case they should fall along the diagonal line, barring a spread owing to detector noise.
In the figure we find this behavior as expected. The measured $\mathcal{M}$ values fall exactly on the diagonal line. This implies that the measured values are highly correlated, which implies the absence of systematics.

The fact that even the presence of the TH terms in the phase of HTF2 waveforms are not able to effect any discriminatory power may not be suprising but is of special significance. While this may be a disappointing result for prospects of characterizing the nature of compact objects in the mass gap, it is important to note that this conclusion is reached with the TH terms in phase. In retrospect, this is not surprising
since Fisher studies point to the same conclusion. These studies, with full waveforms, also indicate that such a distinction is possible in third-generation detectors for binaries within a few hundred Mpc. Bayesian studies with longer waveform simulations for those detectors are computationally expensive and will be pursued elsewhere.

\section{Conclusion}
\label{sec:conclusion}

We have developed a method to search for and characterize  TH in the inspiral phase of a binary. 
We have defined two new parameters that capture the effect of  TH in the inspiral waveform. These parameters are robust enough that even partial absorption can be modeled with them -- something we will pursue in detail in the future. 
To test for the presence of horizon we performed model selection using the Bayes factor. We constructed two sets of waveforms, one for BBHs, which incorporates  TH but no TD, and the other for binaries of horizonless compact objects, which does not include  TH but has TD. We also defined a new effective parameter for the quadrupole moment, namely $\tilde{Q}$, which has been added in both waveform models appropriately.

We showed that for  $m_i>2M_{\odot}$ it was not possible to distinguish between the two models. We did so by employing the Bayes factor in a full Bayesian analysis with simulated injections of both types of signals. We also checked our results with a  Fisher analysis and found that in this mass range it will be hard to test the presence of horizon.

It remains to be seen if it is possible to find better result with golden binaries. It is obvious that with increased SNR the error will decrease, resulting in better measurement of $H_{eff5}$ and $H_{eff8}$. We can estimate the error reduction in such a case. Assuming the sources like in Fig. \ref{fig:horizon_parameters} at $d_L \sim 50$Mpc and taking account of the whole signal duration we can estimate the error reduction. Taking into account of these two, we find that the SNR will increase by a factor of $\sim 3.1$. Hence, the estimation in such case would result in $H_{eff5} \sim 0.06\pm 0.66$ and $H_{eff8} \sim -1.88\pm 7.66$. Hence, with golden binaries even with advanced LIGO detectors, it is possible to find some meaningful constraints on the sources.

An immediate continuation of the current work will be to construct better and more complete waveform models than TidalTF2 and HTF2 that can be used for more precise parameter estimation and more accurate model selection for real signals in contemporaneous GW detector data. Since the waveforms used here were limited to a short part (16s) of the late inspiral phase, it is likely that utilizing more complete waveforms may improve the ability to distinguish BBH and BNSs signals.
Another problem we plan to address is the challenge posed by mixed binaries (NSBH) in discerning the presence of horizons. Thirdly, future generation detectors may allow enough precision so that proper discrimination of NSBH binaries as well as the horizon parameter with intermediate values $(0<H_i<1)$ may be realizable, thereby, affording the possibility of probing the existence of ECOs, such as stellar-mass gravastars, boson stars, etc.~\cite{Datta:2020rvo, Datta:2019epe}.

\section*{Acknowledgments}
We thank Samanwaya Mukherjee for providing useful inputs which helped us express our results better. It is a pleasure to thank  N. V. Krishnendu, Andrea Maselli and Paolo Pani for useful discussions. We would also like to thank Richard Brito and 
Otto Hannuksela for carefully reading the manuscript and providing helpful inputs, and Bhaskar Biswas, Soumak Maitra and Niladri Paul for useful comments. We gratefully acknowledge the use of the IUCAA computing cluster, Sarathi, and the computational resources provided by the LIGO Laboratory (CIT) and supported by National Science Foundation Grants PHY-0757058 and PHY-0823459. SD would like to thank University Grants Commission (UGC), India, for financial support for a senior research fellowship.
KSP acknowledges support of the Netherlands  Organisation  for  Scientific  Research (NWO). 
This work was done with partial  support provided by the Tata Trusts. This paper has been assigned LIGO Document Number LIGO-P2000115. 

{\it We would like to thank all of the essential workers who put their health at their risk during the COVID-19 pandemic, without whom we would not have been able to complete this work.}

\bibliography{main.bib}

\begin{thebibliography}{97}%
\makeatletter
\providecommand \@ifxundefined [1]{%
 \@ifx{#1\undefined}
}%
\providecommand \@ifnum [1]{%
 \ifnum #1\expandafter \@firstoftwo
 \else \expandafter \@secondoftwo
 \fi
}%
\providecommand \@ifx [1]{%
 \ifx #1\expandafter \@firstoftwo
 \else \expandafter \@secondoftwo
 \fi
}%
\providecommand \natexlab [1]{#1}%
\providecommand \enquote  [1]{``#1''}%
\providecommand \bibnamefont  [1]{#1}%
\providecommand \bibfnamefont [1]{#1}%
\providecommand \citenamefont [1]{#1}%
\providecommand \href@noop [0]{\@secondoftwo}%
\providecommand \href [0]{\begingroup \@sanitize@url \@href}%
\providecommand \@href[1]{\@@startlink{#1}\@@href}%
\providecommand \@@href[1]{\endgroup#1\@@endlink}%
\providecommand \@sanitize@url [0]{\catcode `\\12\catcode `\$12\catcode
  `\&12\catcode `\#12\catcode `\^12\catcode `\_12\catcode `\%12\relax}%
\providecommand \@@startlink[1]{}%
\providecommand \@@endlink[0]{}%
\providecommand \url  [0]{\begingroup\@sanitize@url \@url }%
\providecommand \@url [1]{\endgroup\@href {#1}{\urlprefix }}%
\providecommand \urlprefix  [0]{URL }%
\providecommand \Eprint [0]{\href }%
\providecommand \doibase [0]{http://dx.doi.org/}%
\providecommand \selectlanguage [0]{\@gobble}%
\providecommand \bibinfo  [0]{\@secondoftwo}%
\providecommand \bibfield  [0]{\@secondoftwo}%
\providecommand \translation [1]{[#1]}%
\providecommand \BibitemOpen [0]{}%
\providecommand \bibitemStop [0]{}%
\providecommand \bibitemNoStop [0]{.\EOS\space}%
\providecommand \EOS [0]{\spacefactor3000\relax}%
\providecommand \BibitemShut  [1]{\csname bibitem#1\endcsname}%
\let\auto@bib@innerbib\@empty
\bibitem [{\citenamefont {Abbott}\ \emph
  {et~al.}(2019{\natexlab{a}})\citenamefont {Abbott} \emph
  {et~al.}}]{LIGOScientific:2018mvr}%
  \BibitemOpen
  \bibfield  {author} {\bibinfo {author} {\bibfnamefont {B.~P.}\ \bibnamefont
  {Abbott}} \emph {et~al.} (\bibinfo {collaboration} {LIGO Scientific,
  Virgo}),\ }\href {\doibase 10.1103/PhysRevX.9.031040} {\bibfield  {journal}
  {\bibinfo  {journal} {Phys. Rev.}\ }\textbf {\bibinfo {volume} {X9}},\
  \bibinfo {pages} {031040} (\bibinfo {year} {2019}{\natexlab{a}})},\ \Eprint
  {http://arxiv.org/abs/1811.12907} {arXiv:1811.12907 [astro-ph.HE]}
  \BibitemShut {NoStop}%
\bibitem [{\citenamefont {Abbott}\ \emph {et~al.}(2021)\citenamefont {Abbott}
  \emph {et~al.}}]{LIGOScientific:2020ibl}%
  \BibitemOpen
  \bibfield  {author} {\bibinfo {author} {\bibfnamefont {R.}~\bibnamefont
  {Abbott}} \emph {et~al.} (\bibinfo {collaboration} {LIGO Scientific,
  Virgo}),\ }\href {\doibase 10.1103/PhysRevX.11.021053} {\bibfield  {journal}
  {\bibinfo  {journal} {Phys. Rev. X}\ }\textbf {\bibinfo {volume} {11}},\
  \bibinfo {pages} {021053} (\bibinfo {year} {2021})},\ \Eprint
  {http://arxiv.org/abs/2010.14527} {arXiv:2010.14527 [gr-qc]} \BibitemShut
  {NoStop}%
\bibitem [{\citenamefont {Abbott}\ \emph
  {et~al.}(2017{\natexlab{a}})\citenamefont {Abbott} \emph
  {et~al.}}]{gw170817}%
  \BibitemOpen
  \bibfield  {author} {\bibinfo {author} {\bibfnamefont {B.~P.}\ \bibnamefont
  {Abbott}} \emph {et~al.} (\bibinfo {collaboration} {Virgo, LIGO
  Scientific}),\ }\href {\doibase 10.1103/PhysRevLett.119.161101} {\bibfield
  {journal} {\bibinfo  {journal} {Phys. Rev. Lett.}\ }\textbf {\bibinfo
  {volume} {119}},\ \bibinfo {pages} {161101} (\bibinfo {year}
  {2017}{\natexlab{a}})},\ \Eprint {http://arxiv.org/abs/1710.05832}
  {arXiv:1710.05832 [gr-qc]} \BibitemShut {NoStop}%
\bibitem [{\citenamefont {Abbott}\ \emph
  {et~al.}(2019{\natexlab{b}})\citenamefont {Abbott} \emph
  {et~al.}}]{LIGOScientific:2019fpa}%
  \BibitemOpen
  \bibfield  {author} {\bibinfo {author} {\bibfnamefont {B.~P.}\ \bibnamefont
  {Abbott}} \emph {et~al.} (\bibinfo {collaboration} {LIGO Scientific,
  Virgo}),\ }\href {\doibase 10.1103/PhysRevD.100.104036} {\bibfield  {journal}
  {\bibinfo  {journal} {Phys. Rev.}\ }\textbf {\bibinfo {volume} {D100}},\
  \bibinfo {pages} {104036} (\bibinfo {year} {2019}{\natexlab{b}})},\ \Eprint
  {http://arxiv.org/abs/1903.04467} {arXiv:1903.04467 [gr-qc]} \BibitemShut
  {NoStop}%
\bibitem [{\citenamefont {Abbott}\ \emph
  {et~al.}(2019{\natexlab{c}})\citenamefont {Abbott} \emph
  {et~al.}}]{Abbott:2018lct}%
  \BibitemOpen
  \bibfield  {author} {\bibinfo {author} {\bibfnamefont {B.~P.}\ \bibnamefont
  {Abbott}} \emph {et~al.} (\bibinfo {collaboration} {LIGO Scientific,
  Virgo}),\ }\href {\doibase 10.1103/PhysRevLett.123.011102} {\bibfield
  {journal} {\bibinfo  {journal} {Phys. Rev. Lett.}\ }\textbf {\bibinfo
  {volume} {123}},\ \bibinfo {pages} {011102} (\bibinfo {year}
  {2019}{\natexlab{c}})},\ \Eprint {http://arxiv.org/abs/1811.00364}
  {arXiv:1811.00364 [gr-qc]} \BibitemShut {NoStop}%
\bibitem [{\citenamefont {Abbott}\ \emph
  {et~al.}(2016{\natexlab{a}})\citenamefont {Abbott} \emph
  {et~al.}}]{TheLIGOScientific:2016pea}%
  \BibitemOpen
  \bibfield  {author} {\bibinfo {author} {\bibfnamefont {B.~P.}\ \bibnamefont
  {Abbott}} \emph {et~al.} (\bibinfo {collaboration} {LIGO Scientific,
  Virgo}),\ }\href {\doibase 10.1103/PhysRevX.6.041015,
  10.1103/PhysRevX.8.039903} {\bibfield  {journal} {\bibinfo  {journal} {Phys.
  Rev.}\ }\textbf {\bibinfo {volume} {X6}},\ \bibinfo {pages} {041015}
  (\bibinfo {year} {2016}{\natexlab{a}})},\ \bibinfo {note} {[erratum: Phys.
  Rev.X8,no.3,039903(2018)]},\ \Eprint {http://arxiv.org/abs/1606.04856}
  {arXiv:1606.04856 [gr-qc]} \BibitemShut {NoStop}%
\bibitem [{\citenamefont {Abbott}\ \emph
  {et~al.}(2016{\natexlab{b}})\citenamefont {Abbott} \emph
  {et~al.}}]{TheLIGOScientific:2016src}%
  \BibitemOpen
  \bibfield  {author} {\bibinfo {author} {\bibfnamefont {B.~P.}\ \bibnamefont
  {Abbott}} \emph {et~al.} (\bibinfo {collaboration} {LIGO Scientific,
  Virgo}),\ }\href {\doibase 10.1103/PhysRevLett.116.221101,
  10.1103/PhysRevLett.121.129902} {\bibfield  {journal} {\bibinfo  {journal}
  {Phys. Rev. Lett.}\ }\textbf {\bibinfo {volume} {116}},\ \bibinfo {pages}
  {221101} (\bibinfo {year} {2016}{\natexlab{b}})},\ \bibinfo {note} {[Erratum:
  Phys. Rev. Lett.121,no.12,129902(2018)]},\ \Eprint
  {http://arxiv.org/abs/1602.03841} {arXiv:1602.03841 [gr-qc]} \BibitemShut
  {NoStop}%
\bibitem [{\citenamefont {Abbott}\ \emph
  {et~al.}(2017{\natexlab{b}})\citenamefont {Abbott} \emph
  {et~al.}}]{Abbott:2017vtc}%
  \BibitemOpen
  \bibfield  {author} {\bibinfo {author} {\bibfnamefont {B.~P.}\ \bibnamefont
  {Abbott}} \emph {et~al.} (\bibinfo {collaboration} {LIGO Scientific,
  VIRGO}),\ }\href {\doibase 10.1103/PhysRevLett.118.221101,
  10.1103/PhysRevLett.121.129901} {\bibfield  {journal} {\bibinfo  {journal}
  {Phys. Rev. Lett.}\ }\textbf {\bibinfo {volume} {118}},\ \bibinfo {pages}
  {221101} (\bibinfo {year} {2017}{\natexlab{b}})},\ \bibinfo {note} {[Erratum:
  Phys. Rev. Lett.121,no.12,129901(2018)]},\ \Eprint
  {http://arxiv.org/abs/1706.01812} {arXiv:1706.01812 [gr-qc]} \BibitemShut
  {NoStop}%
\bibitem [{\citenamefont {Abbott}\ \emph {et~al.}(2018)\citenamefont {Abbott}
  \emph {et~al.}}]{Abbott:2018exr}%
  \BibitemOpen
  \bibfield  {author} {\bibinfo {author} {\bibfnamefont {B.~P.}\ \bibnamefont
  {Abbott}} \emph {et~al.} (\bibinfo {collaboration} {LIGO Scientific,
  Virgo}),\ }\href {\doibase 10.1103/PhysRevLett.121.161101} {\bibfield
  {journal} {\bibinfo  {journal} {Phys. Rev. Lett.}\ }\textbf {\bibinfo
  {volume} {121}},\ \bibinfo {pages} {161101} (\bibinfo {year} {2018})},\
  \Eprint {http://arxiv.org/abs/1805.11581} {arXiv:1805.11581 [gr-qc]}
  \BibitemShut {NoStop}%
\bibitem [{\citenamefont {De}\ \emph {et~al.}(2018)\citenamefont {De},
  \citenamefont {Finstad}, \citenamefont {Lattimer}, \citenamefont {Brown},
  \citenamefont {Berger},\ and\ \citenamefont {Biwer}}]{De:2018uhw}%
  \BibitemOpen
  \bibfield  {author} {\bibinfo {author} {\bibfnamefont {S.}~\bibnamefont
  {De}}, \bibinfo {author} {\bibfnamefont {D.}~\bibnamefont {Finstad}},
  \bibinfo {author} {\bibfnamefont {J.~M.}\ \bibnamefont {Lattimer}}, \bibinfo
  {author} {\bibfnamefont {D.~A.}\ \bibnamefont {Brown}}, \bibinfo {author}
  {\bibfnamefont {E.}~\bibnamefont {Berger}}, \ and\ \bibinfo {author}
  {\bibfnamefont {C.~M.}\ \bibnamefont {Biwer}},\ }\href {\doibase
  10.1103/PhysRevLett.121.091102} {\bibfield  {journal} {\bibinfo  {journal}
  {Phys. Rev. Lett.}\ }\textbf {\bibinfo {volume} {121}},\ \bibinfo {pages}
  {091102} (\bibinfo {year} {2018})},\ \bibinfo {note} {[Erratum:
  Phys.Rev.Lett. 121, 259902 (2018)]},\ \Eprint
  {http://arxiv.org/abs/1804.08583} {arXiv:1804.08583 [astro-ph.HE]}
  \BibitemShut {NoStop}%
\bibitem [{\citenamefont {Abbott}\ \emph {et~al.}(2020)\citenamefont {Abbott}
  \emph {et~al.}}]{Abbott:2020uma}%
  \BibitemOpen
  \bibfield  {author} {\bibinfo {author} {\bibfnamefont {B.}~\bibnamefont
  {Abbott}} \emph {et~al.} (\bibinfo {collaboration} {LIGO Scientific,
  Virgo}),\ }\href {\doibase 10.3847/2041-8213/ab75f5} {\bibfield  {journal}
  {\bibinfo  {journal} {Astrophys.\ J.\ Lett.}\ }\textbf {\bibinfo {volume}
  {892}},\ \bibinfo {pages} {L3} (\bibinfo {year} {2020})},\ \Eprint
  {http://arxiv.org/abs/2001.01761} {arXiv:2001.01761 [astro-ph.HE]}
  \BibitemShut {NoStop}%
\bibitem [{\citenamefont {Yunes}\ \emph {et~al.}(2016)\citenamefont {Yunes},
  \citenamefont {Yagi},\ and\ \citenamefont {Pretorius}}]{Yunes:2016jcc}%
  \BibitemOpen
  \bibfield  {author} {\bibinfo {author} {\bibfnamefont {N.}~\bibnamefont
  {Yunes}}, \bibinfo {author} {\bibfnamefont {K.}~\bibnamefont {Yagi}}, \ and\
  \bibinfo {author} {\bibfnamefont {F.}~\bibnamefont {Pretorius}},\ }\href
  {\doibase 10.1103/PhysRevD.94.084002} {\bibfield  {journal} {\bibinfo
  {journal} {Phys. Rev.}\ }\textbf {\bibinfo {volume} {D94}},\ \bibinfo {pages}
  {084002} (\bibinfo {year} {2016})},\ \Eprint
  {http://arxiv.org/abs/1603.08955} {arXiv:1603.08955 [gr-qc]} \BibitemShut
  {NoStop}%
\bibitem [{\citenamefont {Cardoso}\ \emph
  {et~al.}(2016{\natexlab{a}})\citenamefont {Cardoso}, \citenamefont {Hopper},
  \citenamefont {Macedo}, \citenamefont {Palenzuela},\ and\ \citenamefont
  {Pani}}]{Cardoso:2016oxy}%
  \BibitemOpen
  \bibfield  {author} {\bibinfo {author} {\bibfnamefont {V.}~\bibnamefont
  {Cardoso}}, \bibinfo {author} {\bibfnamefont {S.}~\bibnamefont {Hopper}},
  \bibinfo {author} {\bibfnamefont {C.~F.~B.}\ \bibnamefont {Macedo}}, \bibinfo
  {author} {\bibfnamefont {C.}~\bibnamefont {Palenzuela}}, \ and\ \bibinfo
  {author} {\bibfnamefont {P.}~\bibnamefont {Pani}},\ }\href {\doibase
  10.1103/PhysRevD.94.084031} {\bibfield  {journal} {\bibinfo  {journal} {Phys.
  Rev.}\ }\textbf {\bibinfo {volume} {D94}},\ \bibinfo {pages} {084031}
  (\bibinfo {year} {2016}{\natexlab{a}})},\ \Eprint
  {http://arxiv.org/abs/1608.08637} {arXiv:1608.08637 [gr-qc]} \BibitemShut
  {NoStop}%
\bibitem [{\citenamefont {Aneesh}\ \emph {et~al.}(2018)\citenamefont {Aneesh},
  \citenamefont {Bose},\ and\ \citenamefont {Kar}}]{Aneesh:2018hlp}%
  \BibitemOpen
  \bibfield  {author} {\bibinfo {author} {\bibfnamefont {S.}~\bibnamefont
  {Aneesh}}, \bibinfo {author} {\bibfnamefont {S.}~\bibnamefont {Bose}}, \ and\
  \bibinfo {author} {\bibfnamefont {S.}~\bibnamefont {Kar}},\ }\href {\doibase
  10.1103/PhysRevD.97.124004} {\bibfield  {journal} {\bibinfo  {journal} {Phys.
  Rev.}\ }\textbf {\bibinfo {volume} {D97}},\ \bibinfo {pages} {124004}
  (\bibinfo {year} {2018})},\ \Eprint {http://arxiv.org/abs/1803.10204}
  {arXiv:1803.10204 [gr-qc]} \BibitemShut {NoStop}%
\bibitem [{\citenamefont {Fryer}\ \emph {et~al.}(2012)\citenamefont {Fryer},
  \citenamefont {Belczynski}, \citenamefont {Wiktorowicz}, \citenamefont
  {Dominik}, \citenamefont {Kalogera},\ and\ \citenamefont
  {Holz}}]{Fryer_2012}%
  \BibitemOpen
  \bibfield  {author} {\bibinfo {author} {\bibfnamefont {C.~L.}\ \bibnamefont
  {Fryer}}, \bibinfo {author} {\bibfnamefont {K.}~\bibnamefont {Belczynski}},
  \bibinfo {author} {\bibfnamefont {G.}~\bibnamefont {Wiktorowicz}}, \bibinfo
  {author} {\bibfnamefont {M.}~\bibnamefont {Dominik}}, \bibinfo {author}
  {\bibfnamefont {V.}~\bibnamefont {Kalogera}}, \ and\ \bibinfo {author}
  {\bibfnamefont {D.~E.}\ \bibnamefont {Holz}},\ }\href {\doibase
  10.1088/0004-637x/749/1/91} {\bibfield  {journal} {\bibinfo  {journal}
  {Astrophys. J.}\ }\textbf {\bibinfo {volume} {749}},\ \bibinfo {pages} {91}
  (\bibinfo {year} {2012})}\BibitemShut {NoStop}%
\bibitem [{\citenamefont {Lunin}\ and\ \citenamefont
  {Mathur}(2002)}]{Lunin:2001jy}%
  \BibitemOpen
  \bibfield  {author} {\bibinfo {author} {\bibfnamefont {O.}~\bibnamefont
  {Lunin}}\ and\ \bibinfo {author} {\bibfnamefont {S.~D.}\ \bibnamefont
  {Mathur}},\ }\href {\doibase 10.1016/S0550-3213(01)00620-4} {\bibfield
  {journal} {\bibinfo  {journal} {Nucl. Phys.}\ }\textbf {\bibinfo {volume}
  {B623}},\ \bibinfo {pages} {342} (\bibinfo {year} {2002})},\ \Eprint
  {http://arxiv.org/abs/hep-th/0109154} {arXiv:hep-th/0109154 [hep-th]}
  \BibitemShut {NoStop}%
\bibitem [{\citenamefont {Almheiri}\ \emph {et~al.}(2013)\citenamefont
  {Almheiri}, \citenamefont {Marolf}, \citenamefont {Polchinski},\ and\
  \citenamefont {Sully}}]{Almheiri:2012rt}%
  \BibitemOpen
  \bibfield  {author} {\bibinfo {author} {\bibfnamefont {A.}~\bibnamefont
  {Almheiri}}, \bibinfo {author} {\bibfnamefont {D.}~\bibnamefont {Marolf}},
  \bibinfo {author} {\bibfnamefont {J.}~\bibnamefont {Polchinski}}, \ and\
  \bibinfo {author} {\bibfnamefont {J.}~\bibnamefont {Sully}},\ }\href
  {\doibase 10.1007/JHEP02(2013)062} {\bibfield  {journal} {\bibinfo  {journal}
  {JHEP}\ }\textbf {\bibinfo {volume} {02}},\ \bibinfo {pages} {062} (\bibinfo
  {year} {2013})},\ \Eprint {http://arxiv.org/abs/1207.3123} {arXiv:1207.3123
  [hep-th]} \BibitemShut {NoStop}%
\bibitem [{\citenamefont {Mazur}\ and\ \citenamefont
  {Mottola}(2004)}]{Mazur:2004fk}%
  \BibitemOpen
  \bibfield  {author} {\bibinfo {author} {\bibfnamefont {P.~O.}\ \bibnamefont
  {Mazur}}\ and\ \bibinfo {author} {\bibfnamefont {E.}~\bibnamefont
  {Mottola}},\ }\href {\doibase 10.1073/pnas.0402717101} {\bibfield  {journal}
  {\bibinfo  {journal} {Proc. Nat. Acad. Sci.}\ }\textbf {\bibinfo {volume}
  {101}},\ \bibinfo {pages} {9545} (\bibinfo {year} {2004})},\ \Eprint
  {http://arxiv.org/abs/gr-qc/0407075} {arXiv:gr-qc/0407075 [gr-qc]}
  \BibitemShut {NoStop}%
\bibitem [{\citenamefont {Liebling}\ and\ \citenamefont
  {Palenzuela}(2012)}]{Liebling:2012fv}%
  \BibitemOpen
  \bibfield  {author} {\bibinfo {author} {\bibfnamefont {S.~L.}\ \bibnamefont
  {Liebling}}\ and\ \bibinfo {author} {\bibfnamefont {C.}~\bibnamefont
  {Palenzuela}},\ }\href {\doibase 10.12942/lrr-2012-6,
  10.1007/s41114-017-0007-y} {\bibfield  {journal} {\bibinfo  {journal} {Living
  Rev. Rel.}\ }\textbf {\bibinfo {volume} {15}},\ \bibinfo {pages} {6}
  (\bibinfo {year} {2012})},\ \bibinfo {note} {[Living Rev.
  Rel.20,no.1,5(2017)]},\ \Eprint {http://arxiv.org/abs/1202.5809}
  {arXiv:1202.5809 [gr-qc]} \BibitemShut {NoStop}%
\bibitem [{\citenamefont {Cardoso}\ \emph
  {et~al.}(2016{\natexlab{b}})\citenamefont {Cardoso}, \citenamefont
  {Franzin},\ and\ \citenamefont {Pani}}]{Cardoso:2016rao}%
  \BibitemOpen
  \bibfield  {author} {\bibinfo {author} {\bibfnamefont {V.}~\bibnamefont
  {Cardoso}}, \bibinfo {author} {\bibfnamefont {E.}~\bibnamefont {Franzin}}, \
  and\ \bibinfo {author} {\bibfnamefont {P.}~\bibnamefont {Pani}},\ }\href
  {\doibase 10.1103/PhysRevLett.116.171101} {\bibfield  {journal} {\bibinfo
  {journal} {Phys. Rev. Lett.}\ }\textbf {\bibinfo {volume} {116}},\ \bibinfo
  {pages} {171101} (\bibinfo {year} {2016}{\natexlab{b}})},\ \bibinfo {note}
  {[Erratum: Phys.Rev.Lett. 117, 089902 (2016)]},\ \Eprint
  {http://arxiv.org/abs/1602.07309} {arXiv:1602.07309 [gr-qc]} \BibitemShut
  {NoStop}%
\bibitem [{\citenamefont {Maggio}\ \emph {et~al.}(2019)\citenamefont {Maggio},
  \citenamefont {Testa}, \citenamefont {Bhagwat},\ and\ \citenamefont
  {Pani}}]{Maggio:2019zyv}%
  \BibitemOpen
  \bibfield  {author} {\bibinfo {author} {\bibfnamefont {E.}~\bibnamefont
  {Maggio}}, \bibinfo {author} {\bibfnamefont {A.}~\bibnamefont {Testa}},
  \bibinfo {author} {\bibfnamefont {S.}~\bibnamefont {Bhagwat}}, \ and\
  \bibinfo {author} {\bibfnamefont {P.}~\bibnamefont {Pani}},\ }\href {\doibase
  10.1103/PhysRevD.100.064056} {\bibfield  {journal} {\bibinfo  {journal}
  {Phys. Rev. D}\ }\textbf {\bibinfo {volume} {100}},\ \bibinfo {pages}
  {064056} (\bibinfo {year} {2019})},\ \Eprint
  {http://arxiv.org/abs/1907.03091} {arXiv:1907.03091 [gr-qc]} \BibitemShut
  {NoStop}%
\bibitem [{\citenamefont {Tsang}\ \emph {et~al.}(2020)\citenamefont {Tsang},
  \citenamefont {Ghosh}, \citenamefont {Samajdar}, \citenamefont
  {Chatziioannou}, \citenamefont {Mastrogiovanni}, \citenamefont {Agathos},\
  and\ \citenamefont {Van Den~Broeck}}]{Tsang:2019zra}%
  \BibitemOpen
  \bibfield  {author} {\bibinfo {author} {\bibfnamefont {K.~W.}\ \bibnamefont
  {Tsang}}, \bibinfo {author} {\bibfnamefont {A.}~\bibnamefont {Ghosh}},
  \bibinfo {author} {\bibfnamefont {A.}~\bibnamefont {Samajdar}}, \bibinfo
  {author} {\bibfnamefont {K.}~\bibnamefont {Chatziioannou}}, \bibinfo {author}
  {\bibfnamefont {S.}~\bibnamefont {Mastrogiovanni}}, \bibinfo {author}
  {\bibfnamefont {M.}~\bibnamefont {Agathos}}, \ and\ \bibinfo {author}
  {\bibfnamefont {C.}~\bibnamefont {Van Den~Broeck}},\ }\href {\doibase
  10.1103/PhysRevD.101.064012} {\bibfield  {journal} {\bibinfo  {journal}
  {Phys.\ Rev.\ D}\ }\textbf {\bibinfo {volume} {101}},\ \bibinfo {pages}
  {064012} (\bibinfo {year} {2020})},\ \Eprint
  {http://arxiv.org/abs/1906.11168} {arXiv:1906.11168 [gr-qc]} \BibitemShut
  {NoStop}%
\bibitem [{\citenamefont {Abedi}\ \emph {et~al.}(2017)\citenamefont {Abedi},
  \citenamefont {Dykaar},\ and\ \citenamefont {Afshordi}}]{Abedi:2016hgu}%
  \BibitemOpen
  \bibfield  {author} {\bibinfo {author} {\bibfnamefont {J.}~\bibnamefont
  {Abedi}}, \bibinfo {author} {\bibfnamefont {H.}~\bibnamefont {Dykaar}}, \
  and\ \bibinfo {author} {\bibfnamefont {N.}~\bibnamefont {Afshordi}},\ }\href
  {\doibase 10.1103/PhysRevD.96.082004} {\bibfield  {journal} {\bibinfo
  {journal} {Phys. Rev.}\ }\textbf {\bibinfo {volume} {D96}},\ \bibinfo {pages}
  {082004} (\bibinfo {year} {2017})},\ \Eprint
  {http://arxiv.org/abs/1612.00266} {arXiv:1612.00266 [gr-qc]} \BibitemShut
  {NoStop}%
\bibitem [{\citenamefont {Westerweck}\ \emph {et~al.}(2018)\citenamefont
  {Westerweck}, \citenamefont {Nielsen}, \citenamefont {Fischer-Birnholtz},
  \citenamefont {Cabero}, \citenamefont {Capano}, \citenamefont {Dent},
  \citenamefont {Krishnan}, \citenamefont {Meadors},\ and\ \citenamefont
  {Nitz}}]{Westerweck:2017hus}%
  \BibitemOpen
  \bibfield  {author} {\bibinfo {author} {\bibfnamefont {J.}~\bibnamefont
  {Westerweck}}, \bibinfo {author} {\bibfnamefont {A.}~\bibnamefont {Nielsen}},
  \bibinfo {author} {\bibfnamefont {O.}~\bibnamefont {Fischer-Birnholtz}},
  \bibinfo {author} {\bibfnamefont {M.}~\bibnamefont {Cabero}}, \bibinfo
  {author} {\bibfnamefont {C.}~\bibnamefont {Capano}}, \bibinfo {author}
  {\bibfnamefont {T.}~\bibnamefont {Dent}}, \bibinfo {author} {\bibfnamefont
  {B.}~\bibnamefont {Krishnan}}, \bibinfo {author} {\bibfnamefont
  {G.}~\bibnamefont {Meadors}}, \ and\ \bibinfo {author} {\bibfnamefont
  {A.~H.}\ \bibnamefont {Nitz}},\ }\href {\doibase 10.1103/PhysRevD.97.124037}
  {\bibfield  {journal} {\bibinfo  {journal} {Phys. Rev.}\ }\textbf {\bibinfo
  {volume} {D97}},\ \bibinfo {pages} {124037} (\bibinfo {year} {2018})},\
  \Eprint {http://arxiv.org/abs/1712.09966} {arXiv:1712.09966 [gr-qc]}
  \BibitemShut {NoStop}%
\bibitem [{\citenamefont {Cardoso}\ and\ \citenamefont
  {Pani}(2019)}]{Cardoso:2019rvt}%
  \BibitemOpen
  \bibfield  {author} {\bibinfo {author} {\bibfnamefont {V.}~\bibnamefont
  {Cardoso}}\ and\ \bibinfo {author} {\bibfnamefont {P.}~\bibnamefont {Pani}},\
  }\href {\doibase 10.1007/s41114-019-0020-4} {\bibfield  {journal} {\bibinfo
  {journal} {Living Rev. Rel.}\ }\textbf {\bibinfo {volume} {22}},\ \bibinfo
  {pages} {4} (\bibinfo {year} {2019})},\ \Eprint
  {http://arxiv.org/abs/1904.05363} {arXiv:1904.05363 [gr-qc]} \BibitemShut
  {NoStop}%
\bibitem [{\citenamefont {Chen}\ \emph {et~al.}(2021)\citenamefont {Chen},
  \citenamefont {Wang},\ and\ \citenamefont {Chen}}]{Chen:2020htz}%
  \BibitemOpen
  \bibfield  {author} {\bibinfo {author} {\bibfnamefont {B.}~\bibnamefont
  {Chen}}, \bibinfo {author} {\bibfnamefont {Q.}~\bibnamefont {Wang}}, \ and\
  \bibinfo {author} {\bibfnamefont {Y.}~\bibnamefont {Chen}},\ }\href {\doibase
  10.1103/PhysRevD.103.104054} {\bibfield  {journal} {\bibinfo  {journal}
  {Phys. Rev. D}\ }\textbf {\bibinfo {volume} {103}},\ \bibinfo {pages}
  {104054} (\bibinfo {year} {2021})},\ \Eprint
  {http://arxiv.org/abs/2012.10842} {arXiv:2012.10842 [gr-qc]} \BibitemShut
  {NoStop}%
\bibitem [{\citenamefont {Xin}\ \emph {et~al.}(2021)\citenamefont {Xin},
  \citenamefont {Chen}, \citenamefont {Lo}, \citenamefont {Sun}, \citenamefont
  {Han}, \citenamefont {Zhong}, \citenamefont {Srivastava}, \citenamefont {Ma},
  \citenamefont {Wang},\ and\ \citenamefont {Chen}}]{Xin:2021zir}%
  \BibitemOpen
  \bibfield  {author} {\bibinfo {author} {\bibfnamefont {S.}~\bibnamefont
  {Xin}}, \bibinfo {author} {\bibfnamefont {B.}~\bibnamefont {Chen}}, \bibinfo
  {author} {\bibfnamefont {R.~K.~L.}\ \bibnamefont {Lo}}, \bibinfo {author}
  {\bibfnamefont {L.}~\bibnamefont {Sun}}, \bibinfo {author} {\bibfnamefont
  {W.-B.}\ \bibnamefont {Han}}, \bibinfo {author} {\bibfnamefont
  {X.}~\bibnamefont {Zhong}}, \bibinfo {author} {\bibfnamefont
  {M.}~\bibnamefont {Srivastava}}, \bibinfo {author} {\bibfnamefont
  {S.}~\bibnamefont {Ma}}, \bibinfo {author} {\bibfnamefont {Q.}~\bibnamefont
  {Wang}}, \ and\ \bibinfo {author} {\bibfnamefont {Y.}~\bibnamefont {Chen}},\
  }\href@noop {} {\  (\bibinfo {year} {2021})},\ \Eprint
  {http://arxiv.org/abs/2105.12313} {arXiv:2105.12313 [gr-qc]} \BibitemShut
  {NoStop}%
\bibitem [{\citenamefont {Cardoso}\ \emph {et~al.}(2017)\citenamefont
  {Cardoso}, \citenamefont {Franzin}, \citenamefont {Maselli}, \citenamefont
  {Pani},\ and\ \citenamefont {Raposo}}]{Cardoso:2017cfl}%
  \BibitemOpen
  \bibfield  {author} {\bibinfo {author} {\bibfnamefont {V.}~\bibnamefont
  {Cardoso}}, \bibinfo {author} {\bibfnamefont {E.}~\bibnamefont {Franzin}},
  \bibinfo {author} {\bibfnamefont {A.}~\bibnamefont {Maselli}}, \bibinfo
  {author} {\bibfnamefont {P.}~\bibnamefont {Pani}}, \ and\ \bibinfo {author}
  {\bibfnamefont {G.}~\bibnamefont {Raposo}},\ }\href {\doibase
  10.1103/PhysRevD.95.089901, 10.1103/PhysRevD.95.084014} {\bibfield  {journal}
  {\bibinfo  {journal} {Phys. Rev.}\ }\textbf {\bibinfo {volume} {D95}},\
  \bibinfo {pages} {084014} (\bibinfo {year} {2017})},\ \bibinfo {note}
  {[Addendum: Phys. Rev.D95,no.8,089901(2017)]},\ \Eprint
  {http://arxiv.org/abs/1701.01116} {arXiv:1701.01116 [gr-qc]} \BibitemShut
  {NoStop}%
\bibitem [{\citenamefont {Sennett}\ \emph {et~al.}(2017)\citenamefont
  {Sennett}, \citenamefont {Hinderer}, \citenamefont {Steinhoff}, \citenamefont
  {Buonanno},\ and\ \citenamefont {Ossokine}}]{Sennett:2017etc}%
  \BibitemOpen
  \bibfield  {author} {\bibinfo {author} {\bibfnamefont {N.}~\bibnamefont
  {Sennett}}, \bibinfo {author} {\bibfnamefont {T.}~\bibnamefont {Hinderer}},
  \bibinfo {author} {\bibfnamefont {J.}~\bibnamefont {Steinhoff}}, \bibinfo
  {author} {\bibfnamefont {A.}~\bibnamefont {Buonanno}}, \ and\ \bibinfo
  {author} {\bibfnamefont {S.}~\bibnamefont {Ossokine}},\ }\href {\doibase
  10.1103/PhysRevD.96.024002} {\bibfield  {journal} {\bibinfo  {journal} {Phys.
  Rev.}\ }\textbf {\bibinfo {volume} {D96}},\ \bibinfo {pages} {024002}
  (\bibinfo {year} {2017})},\ \Eprint {http://arxiv.org/abs/1704.08651}
  {arXiv:1704.08651 [gr-qc]} \BibitemShut {NoStop}%
\bibitem [{\citenamefont {Maselli}\ \emph {et~al.}(2018)\citenamefont
  {Maselli}, \citenamefont {Pani}, \citenamefont {Cardoso}, \citenamefont
  {Abdelsalhin}, \citenamefont {Gualtieri},\ and\ \citenamefont
  {Ferrari}}]{Maselli:2017cmm}%
  \BibitemOpen
  \bibfield  {author} {\bibinfo {author} {\bibfnamefont {A.}~\bibnamefont
  {Maselli}}, \bibinfo {author} {\bibfnamefont {P.}~\bibnamefont {Pani}},
  \bibinfo {author} {\bibfnamefont {V.}~\bibnamefont {Cardoso}}, \bibinfo
  {author} {\bibfnamefont {T.}~\bibnamefont {Abdelsalhin}}, \bibinfo {author}
  {\bibfnamefont {L.}~\bibnamefont {Gualtieri}}, \ and\ \bibinfo {author}
  {\bibfnamefont {V.}~\bibnamefont {Ferrari}},\ }\href {\doibase
  10.1103/PhysRevLett.120.081101} {\bibfield  {journal} {\bibinfo  {journal}
  {Phys. Rev. Lett.}\ }\textbf {\bibinfo {volume} {120}},\ \bibinfo {pages}
  {081101} (\bibinfo {year} {2018})},\ \Eprint
  {http://arxiv.org/abs/1703.10612} {arXiv:1703.10612 [gr-qc]} \BibitemShut
  {NoStop}%
\bibitem [{\citenamefont {Datta}(2021)}]{Datta:2021hvm}%
  \BibitemOpen
  \bibfield  {author} {\bibinfo {author} {\bibfnamefont {S.}~\bibnamefont
  {Datta}},\ }\href@noop {} {\  (\bibinfo {year} {2021})},\ \Eprint
  {http://arxiv.org/abs/2107.07258} {arXiv:2107.07258 [gr-qc]} \BibitemShut
  {NoStop}%
\bibitem [{\citenamefont {Krishnendu}\ \emph {et~al.}(2017)\citenamefont
  {Krishnendu}, \citenamefont {Arun},\ and\ \citenamefont
  {Mishra}}]{Krishnendu:2017shb}%
  \BibitemOpen
  \bibfield  {author} {\bibinfo {author} {\bibfnamefont {N.~V.}\ \bibnamefont
  {Krishnendu}}, \bibinfo {author} {\bibfnamefont {K.~G.}\ \bibnamefont
  {Arun}}, \ and\ \bibinfo {author} {\bibfnamefont {C.~K.}\ \bibnamefont
  {Mishra}},\ }\href {\doibase 10.1103/PhysRevLett.119.091101} {\bibfield
  {journal} {\bibinfo  {journal} {Phys. Rev. Lett.}\ }\textbf {\bibinfo
  {volume} {119}},\ \bibinfo {pages} {091101} (\bibinfo {year} {2017})},\
  \Eprint {http://arxiv.org/abs/1701.06318} {arXiv:1701.06318 [gr-qc]}
  \BibitemShut {NoStop}%
\bibitem [{\citenamefont {Datta}\ and\ \citenamefont
  {Bose}(2019)}]{Datta:2019euh}%
  \BibitemOpen
  \bibfield  {author} {\bibinfo {author} {\bibfnamefont {S.}~\bibnamefont
  {Datta}}\ and\ \bibinfo {author} {\bibfnamefont {S.}~\bibnamefont {Bose}},\
  }\href {\doibase 10.1103/PhysRevD.99.084001} {\bibfield  {journal} {\bibinfo
  {journal} {Phys. Rev.}\ }\textbf {\bibinfo {volume} {D99}},\ \bibinfo {pages}
  {084001} (\bibinfo {year} {2019})},\ \Eprint
  {http://arxiv.org/abs/1902.01723} {arXiv:1902.01723 [gr-qc]} \BibitemShut
  {NoStop}%
\bibitem [{\citenamefont {Bianchi}\ \emph {et~al.}(2020)\citenamefont
  {Bianchi}, \citenamefont {Consoli}, \citenamefont {Grillo}, \citenamefont
  {Morales}, \citenamefont {Pani},\ and\ \citenamefont
  {Raposo}}]{Bianchi:2020bxa}%
  \BibitemOpen
  \bibfield  {author} {\bibinfo {author} {\bibfnamefont {M.}~\bibnamefont
  {Bianchi}}, \bibinfo {author} {\bibfnamefont {D.}~\bibnamefont {Consoli}},
  \bibinfo {author} {\bibfnamefont {A.}~\bibnamefont {Grillo}}, \bibinfo
  {author} {\bibfnamefont {J.~F.}\ \bibnamefont {Morales}}, \bibinfo {author}
  {\bibfnamefont {P.}~\bibnamefont {Pani}}, \ and\ \bibinfo {author}
  {\bibfnamefont {G.}~\bibnamefont {Raposo}},\ }\href {\doibase
  10.1103/PhysRevLett.125.221601} {\bibfield  {journal} {\bibinfo  {journal}
  {Phys. Rev. Lett.}\ }\textbf {\bibinfo {volume} {125}},\ \bibinfo {pages}
  {221601} (\bibinfo {year} {2020})},\ \Eprint
  {http://arxiv.org/abs/2007.01743} {arXiv:2007.01743 [hep-th]} \BibitemShut
  {NoStop}%
\bibitem [{\citenamefont {Mukherjee}\ and\ \citenamefont
  {Chakraborty}(2020)}]{Mukherjee:2020how}%
  \BibitemOpen
  \bibfield  {author} {\bibinfo {author} {\bibfnamefont {S.}~\bibnamefont
  {Mukherjee}}\ and\ \bibinfo {author} {\bibfnamefont {S.}~\bibnamefont
  {Chakraborty}},\ }\href {\doibase 10.1103/PhysRevD.102.124058} {\bibfield
  {journal} {\bibinfo  {journal} {Phys. Rev. D}\ }\textbf {\bibinfo {volume}
  {102}},\ \bibinfo {pages} {124058} (\bibinfo {year} {2020})},\ \Eprint
  {http://arxiv.org/abs/2008.06891} {arXiv:2008.06891 [gr-qc]} \BibitemShut
  {NoStop}%
\bibitem [{\citenamefont {Datta}\ and\ \citenamefont
  {Mukherjee}(2021)}]{Datta:2020axm}%
  \BibitemOpen
  \bibfield  {author} {\bibinfo {author} {\bibfnamefont {S.}~\bibnamefont
  {Datta}}\ and\ \bibinfo {author} {\bibfnamefont {S.}~\bibnamefont
  {Mukherjee}},\ }\href {\doibase 10.1103/PhysRevD.103.104032} {\bibfield
  {journal} {\bibinfo  {journal} {Phys. Rev. D}\ }\textbf {\bibinfo {volume}
  {103}},\ \bibinfo {pages} {104032} (\bibinfo {year} {2021})},\ \Eprint
  {http://arxiv.org/abs/2010.12387} {arXiv:2010.12387 [gr-qc]} \BibitemShut
  {NoStop}%
\bibitem [{\citenamefont {Thorne}\ \emph {et~al.}(1986)\citenamefont {Thorne},
  \citenamefont {Price},\ and\ \citenamefont {Macdonald}}]{MembraneParadigm}%
  \BibitemOpen
  \bibfield  {author} {\bibinfo {author} {\bibfnamefont {K.~S.}\ \bibnamefont
  {Thorne}}, \bibinfo {author} {\bibfnamefont {R.}~\bibnamefont {Price}}, \
  and\ \bibinfo {author} {\bibfnamefont {D.}~\bibnamefont {Macdonald}},\
  }\href@noop {} {\emph {\bibinfo {title} {{Black holes: the membrane
  paradigm}}}},\ edited by\ \bibinfo {editor} {\bibfnamefont {K.~S.}\
  \bibnamefont {Thorne}}\ (\bibinfo  {publisher} {Yale University Press},\
  \bibinfo {year} {1986})\BibitemShut {NoStop}%
\bibitem [{\citenamefont {Damour}(1982)}]{Damour_viscous}%
  \BibitemOpen
  \bibfield  {author} {\bibinfo {author} {\bibfnamefont {T.}~\bibnamefont
  {Damour}},\ }in\ \href@noop {} {\emph {\bibinfo {booktitle} {{Proceedings of
  the Second Marcel Grossmann Meeting ofGeneral Relativity, edited by R.
  Ruffini, North Holland, Amsterdam, 1982 pp 587-608}}}}\ (\bibinfo {year}
  {1982})\BibitemShut {NoStop}%
\bibitem [{\citenamefont {Poisson}(2009)}]{Poisson:2009di}%
  \BibitemOpen
  \bibfield  {author} {\bibinfo {author} {\bibfnamefont {E.}~\bibnamefont
  {Poisson}},\ }\href {\doibase 10.1103/PhysRevD.80.064029} {\bibfield
  {journal} {\bibinfo  {journal} {Phys. Rev.}\ }\textbf {\bibinfo {volume}
  {D80}},\ \bibinfo {pages} {064029} (\bibinfo {year} {2009})},\ \Eprint
  {http://arxiv.org/abs/0907.0874} {arXiv:0907.0874 [gr-qc]} \BibitemShut
  {NoStop}%
\bibitem [{\citenamefont {Cardoso}\ and\ \citenamefont
  {Pani}(2013)}]{Cardoso:2012zn}%
  \BibitemOpen
  \bibfield  {author} {\bibinfo {author} {\bibfnamefont {V.}~\bibnamefont
  {Cardoso}}\ and\ \bibinfo {author} {\bibfnamefont {P.}~\bibnamefont {Pani}},\
  }\href {\doibase 10.1088/0264-9381/30/4/045011} {\bibfield  {journal}
  {\bibinfo  {journal} {Class. Quant. Grav.}\ }\textbf {\bibinfo {volume}
  {30}},\ \bibinfo {pages} {045011} (\bibinfo {year} {2013})},\ \Eprint
  {http://arxiv.org/abs/1205.3184} {arXiv:1205.3184 [gr-qc]} \BibitemShut
  {NoStop}%
\bibitem [{\citenamefont {Hartle}(1973)}]{Hartle:1973zz}%
  \BibitemOpen
  \bibfield  {author} {\bibinfo {author} {\bibfnamefont {J.~B.}\ \bibnamefont
  {Hartle}},\ }\href {\doibase 10.1103/PhysRevD.8.1010} {\bibfield  {journal}
  {\bibinfo  {journal} {Phys. Rev.}\ }\textbf {\bibinfo {volume} {D8}},\
  \bibinfo {pages} {1010} (\bibinfo {year} {1973})}\BibitemShut {NoStop}%
\bibitem [{\citenamefont {Hughes}(2001)}]{Hughes:2001jr}%
  \BibitemOpen
  \bibfield  {author} {\bibinfo {author} {\bibfnamefont {S.~A.}\ \bibnamefont
  {Hughes}},\ }\href {\doibase 10.1103/PhysRevD.64.064004,
  10.1103/PhysRevD.88.109902} {\bibfield  {journal} {\bibinfo  {journal} {Phys.
  Rev.}\ }\textbf {\bibinfo {volume} {D64}},\ \bibinfo {pages} {064004}
  (\bibinfo {year} {2001})},\ \bibinfo {note} {[Erratum: Phys.
  Rev.D88,no.10,109902(2013)]},\ \Eprint {http://arxiv.org/abs/gr-qc/0104041}
  {arXiv:gr-qc/0104041 [gr-qc]} \BibitemShut {NoStop}%
\bibitem [{\citenamefont {Poisson}\ and\ \citenamefont
  {Will}(1953)}]{PoissonWill}%
  \BibitemOpen
  \bibfield  {author} {\bibinfo {author} {\bibfnamefont {E.}~\bibnamefont
  {Poisson}}\ and\ \bibinfo {author} {\bibfnamefont {C.}~\bibnamefont {Will}},\
  }\href@noop {} {\emph {\bibinfo {title} {{Gravity: Newtonian, Post-Newtonian,
  Relativistic}}}}\ (\bibinfo  {publisher} {Cambridge University Press},\
  \bibinfo {address} {Cambridge, UK},\ \bibinfo {year} {1953})\BibitemShut
  {NoStop}%
\bibitem [{\citenamefont {Chakraborty}\ \emph {et~al.}(2021)\citenamefont
  {Chakraborty}, \citenamefont {Datta},\ and\ \citenamefont
  {Sau}}]{Chakraborty:2021gdf}%
  \BibitemOpen
  \bibfield  {author} {\bibinfo {author} {\bibfnamefont {S.}~\bibnamefont
  {Chakraborty}}, \bibinfo {author} {\bibfnamefont {S.}~\bibnamefont {Datta}},
  \ and\ \bibinfo {author} {\bibfnamefont {S.}~\bibnamefont {Sau}},\
  }\href@noop {} {\  (\bibinfo {year} {2021})},\ \Eprint
  {http://arxiv.org/abs/2103.12430} {arXiv:2103.12430 [gr-qc]} \BibitemShut
  {NoStop}%
\bibitem [{\citenamefont {Datta}(2020)}]{Datta:2020rvo}%
  \BibitemOpen
  \bibfield  {author} {\bibinfo {author} {\bibfnamefont {S.}~\bibnamefont
  {Datta}},\ }\href {\doibase 10.1103/PhysRevD.102.064040} {\bibfield
  {journal} {\bibinfo  {journal} {Phys. Rev. D}\ }\textbf {\bibinfo {volume}
  {102}},\ \bibinfo {pages} {064040} (\bibinfo {year} {2020})},\ \Eprint
  {http://arxiv.org/abs/2002.04480} {arXiv:2002.04480 [gr-qc]} \BibitemShut
  {NoStop}%
\bibitem [{\citenamefont {Agullo}\ \emph {et~al.}(2021)\citenamefont {Agullo},
  \citenamefont {Cardoso}, \citenamefont {Rio}, \citenamefont {Maggiore},\ and\
  \citenamefont {Pullin}}]{Agullo:2020hxe}%
  \BibitemOpen
  \bibfield  {author} {\bibinfo {author} {\bibfnamefont {I.}~\bibnamefont
  {Agullo}}, \bibinfo {author} {\bibfnamefont {V.}~\bibnamefont {Cardoso}},
  \bibinfo {author} {\bibfnamefont {A.~D.}\ \bibnamefont {Rio}}, \bibinfo
  {author} {\bibfnamefont {M.}~\bibnamefont {Maggiore}}, \ and\ \bibinfo
  {author} {\bibfnamefont {J.}~\bibnamefont {Pullin}},\ }\href {\doibase
  10.1103/PhysRevLett.126.041302} {\bibfield  {journal} {\bibinfo  {journal}
  {Phys. Rev. Lett.}\ }\textbf {\bibinfo {volume} {126}},\ \bibinfo {pages}
  {041302} (\bibinfo {year} {2021})},\ \Eprint
  {http://arxiv.org/abs/2007.13761} {arXiv:2007.13761 [gr-qc]} \BibitemShut
  {NoStop}%
\bibitem [{\citenamefont {Datta}\ and\ \citenamefont
  {Phukon}(2021)}]{Datta:2021row}%
  \BibitemOpen
  \bibfield  {author} {\bibinfo {author} {\bibfnamefont {S.}~\bibnamefont
  {Datta}}\ and\ \bibinfo {author} {\bibfnamefont {K.~S.}\ \bibnamefont
  {Phukon}},\ }\href@noop {} {\  (\bibinfo {year} {2021})},\ \Eprint
  {http://arxiv.org/abs/2105.11140} {arXiv:2105.11140 [gr-qc]} \BibitemShut
  {NoStop}%
\bibitem [{\citenamefont {Sago}\ and\ \citenamefont
  {Tanaka}(2021)}]{Sago:2021iku}%
  \BibitemOpen
  \bibfield  {author} {\bibinfo {author} {\bibfnamefont {N.}~\bibnamefont
  {Sago}}\ and\ \bibinfo {author} {\bibfnamefont {T.}~\bibnamefont {Tanaka}},\
  }\href@noop {} {\  (\bibinfo {year} {2021})},\ \Eprint
  {http://arxiv.org/abs/2106.07123} {arXiv:2106.07123 [gr-qc]} \BibitemShut
  {NoStop}%
\bibitem [{\citenamefont {Datta}\ \emph {et~al.}(2020)\citenamefont {Datta},
  \citenamefont {Brito}, \citenamefont {Bose}, \citenamefont {Pani},\ and\
  \citenamefont {Hughes}}]{Datta:2019epe}%
  \BibitemOpen
  \bibfield  {author} {\bibinfo {author} {\bibfnamefont {S.}~\bibnamefont
  {Datta}}, \bibinfo {author} {\bibfnamefont {R.}~\bibnamefont {Brito}},
  \bibinfo {author} {\bibfnamefont {S.}~\bibnamefont {Bose}}, \bibinfo {author}
  {\bibfnamefont {P.}~\bibnamefont {Pani}}, \ and\ \bibinfo {author}
  {\bibfnamefont {S.~A.}\ \bibnamefont {Hughes}},\ }\href {\doibase
  10.1103/PhysRevD.101.044004} {\bibfield  {journal} {\bibinfo  {journal}
  {Phys. Rev.}\ }\textbf {\bibinfo {volume} {D101}},\ \bibinfo {pages} {044004}
  (\bibinfo {year} {2020})},\ \Eprint {http://arxiv.org/abs/1910.07841}
  {arXiv:1910.07841 [gr-qc]} \BibitemShut {NoStop}%
\bibitem [{\citenamefont {Akutsu}\ \emph {et~al.}(2020)\citenamefont {Akutsu}
  \emph {et~al.}}]{Akutsu:2020zlw}%
  \BibitemOpen
  \bibfield  {author} {\bibinfo {author} {\bibfnamefont {T.}~\bibnamefont
  {Akutsu}} \emph {et~al.} (\bibinfo {collaboration} {KAGRA}),\ }\href
  {\doibase 10.1093/ptep/ptaa120} {\bibfield  {journal} {\bibinfo  {journal}
  {Progress of Theoretical and Experimental Physics}\ } (\bibinfo {year}
  {2020}),\ 10.1093/ptep/ptaa120},\ \Eprint {http://arxiv.org/abs/2008.02921}
  {arXiv:2008.02921 [gr-qc]} \BibitemShut {NoStop}%
\bibitem [{\citenamefont {Glampedakis}\ \emph {et~al.}(2014)\citenamefont
  {Glampedakis}, \citenamefont {Kapadia},\ and\ \citenamefont
  {Kennefick}}]{Glampedakis:2013jya}%
  \BibitemOpen
  \bibfield  {author} {\bibinfo {author} {\bibfnamefont {K.}~\bibnamefont
  {Glampedakis}}, \bibinfo {author} {\bibfnamefont {S.~J.}\ \bibnamefont
  {Kapadia}}, \ and\ \bibinfo {author} {\bibfnamefont {D.}~\bibnamefont
  {Kennefick}},\ }\href {\doibase 10.1103/PhysRevD.89.024007} {\bibfield
  {journal} {\bibinfo  {journal} {Phys. Rev.}\ }\textbf {\bibinfo {volume}
  {D89}},\ \bibinfo {pages} {024007} (\bibinfo {year} {2014})},\ \Eprint
  {http://arxiv.org/abs/1312.1912} {arXiv:1312.1912 [gr-qc]} \BibitemShut
  {NoStop}%
\bibitem [{\citenamefont {Flanagan}\ and\ \citenamefont
  {Hinderer}(2008)}]{Flanagan:2007ix}%
  \BibitemOpen
  \bibfield  {author} {\bibinfo {author} {\bibfnamefont {E.~E.}\ \bibnamefont
  {Flanagan}}\ and\ \bibinfo {author} {\bibfnamefont {T.}~\bibnamefont
  {Hinderer}},\ }\href {\doibase 10.1103/PhysRevD.77.021502} {\bibfield
  {journal} {\bibinfo  {journal} {Phys. Rev. D}\ }\textbf {\bibinfo {volume}
  {77}},\ \bibinfo {pages} {021502} (\bibinfo {year} {2008})},\ \Eprint
  {http://arxiv.org/abs/0709.1915} {arXiv:0709.1915 [astro-ph]} \BibitemShut
  {NoStop}%
\bibitem [{\citenamefont {Damour}\ and\ \citenamefont
  {Nagar}(2009)}]{Damour:2009vw}%
  \BibitemOpen
  \bibfield  {author} {\bibinfo {author} {\bibfnamefont {T.}~\bibnamefont
  {Damour}}\ and\ \bibinfo {author} {\bibfnamefont {A.}~\bibnamefont {Nagar}},\
  }\href {\doibase 10.1103/PhysRevD.80.084035} {\bibfield  {journal} {\bibinfo
  {journal} {Phys. Rev.}\ }\textbf {\bibinfo {volume} {D80}},\ \bibinfo {pages}
  {084035} (\bibinfo {year} {2009})},\ \Eprint {http://arxiv.org/abs/0906.0096}
  {arXiv:0906.0096 [gr-qc]} \BibitemShut {NoStop}%
\bibitem [{\citenamefont {Binnington}\ and\ \citenamefont
  {Poisson}(2009)}]{Binnington:2009bb}%
  \BibitemOpen
  \bibfield  {author} {\bibinfo {author} {\bibfnamefont {T.}~\bibnamefont
  {Binnington}}\ and\ \bibinfo {author} {\bibfnamefont {E.}~\bibnamefont
  {Poisson}},\ }\href {\doibase 10.1103/PhysRevD.80.084018} {\bibfield
  {journal} {\bibinfo  {journal} {Phys. Rev.}\ }\textbf {\bibinfo {volume}
  {D80}},\ \bibinfo {pages} {084018} (\bibinfo {year} {2009})},\ \Eprint
  {http://arxiv.org/abs/0906.1366} {arXiv:0906.1366 [gr-qc]} \BibitemShut
  {NoStop}%
\bibitem [{\citenamefont {Chakravarti}\ \emph {et~al.}(2019)\citenamefont
  {Chakravarti}, \citenamefont {Chakraborty}, \citenamefont {Bose},\ and\
  \citenamefont {SenGupta}}]{Chakravarti:2018vlt}%
  \BibitemOpen
  \bibfield  {author} {\bibinfo {author} {\bibfnamefont {K.}~\bibnamefont
  {Chakravarti}}, \bibinfo {author} {\bibfnamefont {S.}~\bibnamefont
  {Chakraborty}}, \bibinfo {author} {\bibfnamefont {S.}~\bibnamefont {Bose}}, \
  and\ \bibinfo {author} {\bibfnamefont {S.}~\bibnamefont {SenGupta}},\ }\href
  {\doibase 10.1103/PhysRevD.99.024036} {\bibfield  {journal} {\bibinfo
  {journal} {Phys. Rev.}\ }\textbf {\bibinfo {volume} {D99}},\ \bibinfo {pages}
  {024036} (\bibinfo {year} {2019})},\ \Eprint
  {http://arxiv.org/abs/1811.11364} {arXiv:1811.11364 [gr-qc]} \BibitemShut
  {NoStop}%
\bibitem [{\citenamefont {Brustein}\ and\ \citenamefont
  {Sherf}(2020)}]{Brustein:2020tpg}%
  \BibitemOpen
  \bibfield  {author} {\bibinfo {author} {\bibfnamefont {R.}~\bibnamefont
  {Brustein}}\ and\ \bibinfo {author} {\bibfnamefont {Y.}~\bibnamefont
  {Sherf}},\ }\href@noop {} {\  (\bibinfo {year} {2020})},\ \Eprint
  {http://arxiv.org/abs/2008.02738} {arXiv:2008.02738 [gr-qc]} \BibitemShut
  {NoStop}%
\bibitem [{\citenamefont {Landry}\ and\ \citenamefont
  {Poisson}(2014)}]{Landry:2014jka}%
  \BibitemOpen
  \bibfield  {author} {\bibinfo {author} {\bibfnamefont {P.}~\bibnamefont
  {Landry}}\ and\ \bibinfo {author} {\bibfnamefont {E.}~\bibnamefont
  {Poisson}},\ }\href {\doibase 10.1103/PhysRevD.89.124011} {\bibfield
  {journal} {\bibinfo  {journal} {Phys. Rev. D}\ }\textbf {\bibinfo {volume}
  {89}},\ \bibinfo {pages} {124011} (\bibinfo {year} {2014})},\ \Eprint
  {http://arxiv.org/abs/1404.6798} {arXiv:1404.6798 [gr-qc]} \BibitemShut
  {NoStop}%
\bibitem [{\citenamefont {Chia}(2020)}]{Chia:2020yla}%
  \BibitemOpen
  \bibfield  {author} {\bibinfo {author} {\bibfnamefont {H.~S.}\ \bibnamefont
  {Chia}},\ }\href@noop {} {\  (\bibinfo {year} {2020})},\ \Eprint
  {http://arxiv.org/abs/2010.07300} {arXiv:2010.07300 [gr-qc]} \BibitemShut
  {NoStop}%
\bibitem [{\citenamefont {Le~Tiec}\ \emph {et~al.}(2021)\citenamefont
  {Le~Tiec}, \citenamefont {Casals},\ and\ \citenamefont
  {Franzin}}]{LeTiec:2020bos}%
  \BibitemOpen
  \bibfield  {author} {\bibinfo {author} {\bibfnamefont {A.}~\bibnamefont
  {Le~Tiec}}, \bibinfo {author} {\bibfnamefont {M.}~\bibnamefont {Casals}}, \
  and\ \bibinfo {author} {\bibfnamefont {E.}~\bibnamefont {Franzin}},\ }\href
  {\doibase 10.1103/PhysRevD.103.084021} {\bibfield  {journal} {\bibinfo
  {journal} {Phys. Rev. D}\ }\textbf {\bibinfo {volume} {103}},\ \bibinfo
  {pages} {084021} (\bibinfo {year} {2021})},\ \Eprint
  {http://arxiv.org/abs/2010.15795} {arXiv:2010.15795 [gr-qc]} \BibitemShut
  {NoStop}%
\bibitem [{\citenamefont {Blanchet}(2014)}]{Blanchet:2013haa}%
  \BibitemOpen
  \bibfield  {author} {\bibinfo {author} {\bibfnamefont {L.}~\bibnamefont
  {Blanchet}},\ }\href {\doibase 10.12942/lrr-2014-2} {\bibfield  {journal}
  {\bibinfo  {journal} {Living Rev. Rel.}\ }\textbf {\bibinfo {volume} {17}},\
  \bibinfo {pages} {2} (\bibinfo {year} {2014})},\ \Eprint
  {http://arxiv.org/abs/1310.1528} {arXiv:1310.1528 [gr-qc]} \BibitemShut
  {NoStop}%
\bibitem [{\citenamefont {Alvi}(2001)}]{Alvi:2001mx}%
  \BibitemOpen
  \bibfield  {author} {\bibinfo {author} {\bibfnamefont {K.}~\bibnamefont
  {Alvi}},\ }\href {\doibase 10.1103/PhysRevD.64.104020} {\bibfield  {journal}
  {\bibinfo  {journal} {Phys. Rev.}\ }\textbf {\bibinfo {volume} {D64}},\
  \bibinfo {pages} {104020} (\bibinfo {year} {2001})},\ \Eprint
  {http://arxiv.org/abs/gr-qc/0107080} {arXiv:gr-qc/0107080 [gr-qc]}
  \BibitemShut {NoStop}%
\bibitem [{\citenamefont {Poisson}\ and\ \citenamefont
  {Corrigan}(2018)}]{Poisson:2018qqd}%
  \BibitemOpen
  \bibfield  {author} {\bibinfo {author} {\bibfnamefont {E.}~\bibnamefont
  {Poisson}}\ and\ \bibinfo {author} {\bibfnamefont {E.}~\bibnamefont
  {Corrigan}},\ }\href {\doibase 10.1103/PhysRevD.97.124048} {\bibfield
  {journal} {\bibinfo  {journal} {Phys. Rev.}\ }\textbf {\bibinfo {volume}
  {D97}},\ \bibinfo {pages} {124048} (\bibinfo {year} {2018})},\ \Eprint
  {http://arxiv.org/abs/1804.01848} {arXiv:1804.01848 [gr-qc]} \BibitemShut
  {NoStop}%
\bibitem [{\citenamefont {Nagar}\ and\ \citenamefont
  {Akcay}(2012)}]{Nagar:2011aa}%
  \BibitemOpen
  \bibfield  {author} {\bibinfo {author} {\bibfnamefont {A.}~\bibnamefont
  {Nagar}}\ and\ \bibinfo {author} {\bibfnamefont {S.}~\bibnamefont {Akcay}},\
  }\href {\doibase 10.1103/PhysRevD.85.044025} {\bibfield  {journal} {\bibinfo
  {journal} {Phys. Rev.}\ }\textbf {\bibinfo {volume} {D85}},\ \bibinfo {pages}
  {044025} (\bibinfo {year} {2012})},\ \Eprint {http://arxiv.org/abs/1112.2840}
  {arXiv:1112.2840 [gr-qc]} \BibitemShut {NoStop}%
\bibitem [{\citenamefont {Bernuzzi}\ \emph {et~al.}(2012)\citenamefont
  {Bernuzzi}, \citenamefont {Nagar},\ and\ \citenamefont
  {Zenginoglu}}]{Bernuzzi:2012ku}%
  \BibitemOpen
  \bibfield  {author} {\bibinfo {author} {\bibfnamefont {S.}~\bibnamefont
  {Bernuzzi}}, \bibinfo {author} {\bibfnamefont {A.}~\bibnamefont {Nagar}}, \
  and\ \bibinfo {author} {\bibfnamefont {A.}~\bibnamefont {Zenginoglu}},\
  }\href {\doibase 10.1103/PhysRevD.86.104038} {\bibfield  {journal} {\bibinfo
  {journal} {Phys. Rev.}\ }\textbf {\bibinfo {volume} {D86}},\ \bibinfo {pages}
  {104038} (\bibinfo {year} {2012})},\ \Eprint {http://arxiv.org/abs/1207.0769}
  {arXiv:1207.0769 [gr-qc]} \BibitemShut {NoStop}%
\bibitem [{\citenamefont {Chatziioannou}\ \emph {et~al.}(2016)\citenamefont
  {Chatziioannou}, \citenamefont {Poisson},\ and\ \citenamefont
  {Yunes}}]{Chatziioannou:2016kem}%
  \BibitemOpen
  \bibfield  {author} {\bibinfo {author} {\bibfnamefont {K.}~\bibnamefont
  {Chatziioannou}}, \bibinfo {author} {\bibfnamefont {E.}~\bibnamefont
  {Poisson}}, \ and\ \bibinfo {author} {\bibfnamefont {N.}~\bibnamefont
  {Yunes}},\ }\href {\doibase 10.1103/PhysRevD.94.084043} {\bibfield  {journal}
  {\bibinfo  {journal} {Phys. Rev.}\ }\textbf {\bibinfo {volume} {D94}},\
  \bibinfo {pages} {084043} (\bibinfo {year} {2016})},\ \Eprint
  {http://arxiv.org/abs/1608.02899} {arXiv:1608.02899 [gr-qc]} \BibitemShut
  {NoStop}%
\bibitem [{\citenamefont {Damour}(2001)}]{Damour:2001tu}%
  \BibitemOpen
  \bibfield  {author} {\bibinfo {author} {\bibfnamefont {T.}~\bibnamefont
  {Damour}},\ }\href {\doibase 10.1103/PhysRevD.64.124013} {\bibfield
  {journal} {\bibinfo  {journal} {Phys. Rev.}\ }\textbf {\bibinfo {volume}
  {D64}},\ \bibinfo {pages} {124013} (\bibinfo {year} {2001})},\ \Eprint
  {http://arxiv.org/abs/gr-qc/0103018} {arXiv:gr-qc/0103018 [gr-qc]}
  \BibitemShut {NoStop}%
\bibitem [{\citenamefont {Racine}(2008)}]{Racine:2008qv}%
  \BibitemOpen
  \bibfield  {author} {\bibinfo {author} {\bibfnamefont {E.}~\bibnamefont
  {Racine}},\ }\href {\doibase 10.1103/PhysRevD.78.044021} {\bibfield
  {journal} {\bibinfo  {journal} {Phys. Rev.}\ }\textbf {\bibinfo {volume}
  {D78}},\ \bibinfo {pages} {044021} (\bibinfo {year} {2008})},\ \Eprint
  {http://arxiv.org/abs/0803.1820} {arXiv:0803.1820 [gr-qc]} \BibitemShut
  {NoStop}%
\bibitem [{\citenamefont {Ajith}(2011)}]{Ajith:2011ec}%
  \BibitemOpen
  \bibfield  {author} {\bibinfo {author} {\bibfnamefont {P.}~\bibnamefont
  {Ajith}},\ }\href {\doibase 10.1103/PhysRevD.84.084037} {\bibfield  {journal}
  {\bibinfo  {journal} {Phys. Rev.}\ }\textbf {\bibinfo {volume} {D84}},\
  \bibinfo {pages} {084037} (\bibinfo {year} {2011})},\ \Eprint
  {http://arxiv.org/abs/1107.1267} {arXiv:1107.1267 [gr-qc]} \BibitemShut
  {NoStop}%
\bibitem [{\citenamefont {Cutler}\ and\ \citenamefont
  {Flanagan}(1994)}]{Cutler:1994ys}%
  \BibitemOpen
  \bibfield  {author} {\bibinfo {author} {\bibfnamefont {C.}~\bibnamefont
  {Cutler}}\ and\ \bibinfo {author} {\bibfnamefont {E.~E.}\ \bibnamefont
  {Flanagan}},\ }\href {\doibase 10.1103/PhysRevD.49.2658} {\bibfield
  {journal} {\bibinfo  {journal} {Phys. Rev.}\ }\textbf {\bibinfo {volume}
  {D49}},\ \bibinfo {pages} {2658} (\bibinfo {year} {1994})},\ \Eprint
  {http://arxiv.org/abs/gr-qc/9402014} {arXiv:gr-qc/9402014 [gr-qc]}
  \BibitemShut {NoStop}%
\bibitem [{\citenamefont {Husa}\ \emph {et~al.}(2016)\citenamefont {Husa},
  \citenamefont {Khan}, \citenamefont {Hannam}, \citenamefont {Pürrer},
  \citenamefont {Ohme}, \citenamefont {Jiménez~Forteza},\ and\ \citenamefont
  {Bohé}}]{Husa:2015iqa}%
  \BibitemOpen
  \bibfield  {author} {\bibinfo {author} {\bibfnamefont {S.}~\bibnamefont
  {Husa}}, \bibinfo {author} {\bibfnamefont {S.}~\bibnamefont {Khan}}, \bibinfo
  {author} {\bibfnamefont {M.}~\bibnamefont {Hannam}}, \bibinfo {author}
  {\bibfnamefont {M.}~\bibnamefont {Pürrer}}, \bibinfo {author} {\bibfnamefont
  {F.}~\bibnamefont {Ohme}}, \bibinfo {author} {\bibfnamefont {X.}~\bibnamefont
  {Jiménez~Forteza}}, \ and\ \bibinfo {author} {\bibfnamefont
  {A.}~\bibnamefont {Bohé}},\ }\href {\doibase 10.1103/PhysRevD.93.044006}
  {\bibfield  {journal} {\bibinfo  {journal} {Phys. Rev.}\ }\textbf {\bibinfo
  {volume} {D93}},\ \bibinfo {pages} {044006} (\bibinfo {year} {2016})},\
  \Eprint {http://arxiv.org/abs/1508.07250} {arXiv:1508.07250 [gr-qc]}
  \BibitemShut {NoStop}%
\bibitem [{\citenamefont {Khan}\ \emph {et~al.}(2016)\citenamefont {Khan},
  \citenamefont {Husa}, \citenamefont {Hannam}, \citenamefont {Ohme},
  \citenamefont {Pürrer}, \citenamefont {Jiménez~Forteza},\ and\
  \citenamefont {Bohé}}]{Khan:2015jqa}%
  \BibitemOpen
  \bibfield  {author} {\bibinfo {author} {\bibfnamefont {S.}~\bibnamefont
  {Khan}}, \bibinfo {author} {\bibfnamefont {S.}~\bibnamefont {Husa}}, \bibinfo
  {author} {\bibfnamefont {M.}~\bibnamefont {Hannam}}, \bibinfo {author}
  {\bibfnamefont {F.}~\bibnamefont {Ohme}}, \bibinfo {author} {\bibfnamefont
  {M.}~\bibnamefont {Pürrer}}, \bibinfo {author} {\bibfnamefont
  {X.}~\bibnamefont {Jiménez~Forteza}}, \ and\ \bibinfo {author}
  {\bibfnamefont {A.}~\bibnamefont {Bohé}},\ }\href {\doibase
  10.1103/PhysRevD.93.044007} {\bibfield  {journal} {\bibinfo  {journal} {Phys.
  Rev.}\ }\textbf {\bibinfo {volume} {D93}},\ \bibinfo {pages} {044007}
  (\bibinfo {year} {2016})},\ \Eprint {http://arxiv.org/abs/1508.07253}
  {arXiv:1508.07253 [gr-qc]} \BibitemShut {NoStop}%
\bibitem [{\citenamefont {Hannam}\ \emph {et~al.}(2014)\citenamefont {Hannam},
  \citenamefont {Schmidt}, \citenamefont {Bohé}, \citenamefont {Haegel},
  \citenamefont {Husa}, \citenamefont {Ohme}, \citenamefont {Pratten},\ and\
  \citenamefont {Pürrer}}]{Hannam:2013oca}%
  \BibitemOpen
  \bibfield  {author} {\bibinfo {author} {\bibfnamefont {M.}~\bibnamefont
  {Hannam}}, \bibinfo {author} {\bibfnamefont {P.}~\bibnamefont {Schmidt}},
  \bibinfo {author} {\bibfnamefont {A.}~\bibnamefont {Bohé}}, \bibinfo
  {author} {\bibfnamefont {L.}~\bibnamefont {Haegel}}, \bibinfo {author}
  {\bibfnamefont {S.}~\bibnamefont {Husa}}, \bibinfo {author} {\bibfnamefont
  {F.}~\bibnamefont {Ohme}}, \bibinfo {author} {\bibfnamefont {G.}~\bibnamefont
  {Pratten}}, \ and\ \bibinfo {author} {\bibfnamefont {M.}~\bibnamefont
  {Pürrer}},\ }\href {\doibase 10.1103/PhysRevLett.113.151101} {\bibfield
  {journal} {\bibinfo  {journal} {Phys. Rev. Lett.}\ }\textbf {\bibinfo
  {volume} {113}},\ \bibinfo {pages} {151101} (\bibinfo {year} {2014})},\
  \Eprint {http://arxiv.org/abs/1308.3271} {arXiv:1308.3271 [gr-qc]}
  \BibitemShut {NoStop}%
\bibitem [{\citenamefont {Tichy}\ \emph {et~al.}(2000)\citenamefont {Tichy},
  \citenamefont {Flanagan},\ and\ \citenamefont {Poisson}}]{Tichy:1999pv}%
  \BibitemOpen
  \bibfield  {author} {\bibinfo {author} {\bibfnamefont {W.}~\bibnamefont
  {Tichy}}, \bibinfo {author} {\bibfnamefont {E.~E.}\ \bibnamefont {Flanagan}},
  \ and\ \bibinfo {author} {\bibfnamefont {E.}~\bibnamefont {Poisson}},\ }\href
  {\doibase 10.1103/PhysRevD.61.104015} {\bibfield  {journal} {\bibinfo
  {journal} {Phys. Rev.}\ }\textbf {\bibinfo {volume} {D61}},\ \bibinfo {pages}
  {104015} (\bibinfo {year} {2000})},\ \Eprint
  {http://arxiv.org/abs/gr-qc/9912075} {arXiv:gr-qc/9912075 [gr-qc]}
  \BibitemShut {NoStop}%
\bibitem [{\citenamefont {Isoyama}\ and\ \citenamefont
  {Nakano}(2018)}]{Isoyama:2017tbp}%
  \BibitemOpen
  \bibfield  {author} {\bibinfo {author} {\bibfnamefont {S.}~\bibnamefont
  {Isoyama}}\ and\ \bibinfo {author} {\bibfnamefont {H.}~\bibnamefont
  {Nakano}},\ }\href {\doibase 10.1088/1361-6382/aa96c5} {\bibfield  {journal}
  {\bibinfo  {journal} {Class. Quant. Grav.}\ }\textbf {\bibinfo {volume}
  {35}},\ \bibinfo {pages} {024001} (\bibinfo {year} {2018})},\ \Eprint
  {http://arxiv.org/abs/1705.03869} {arXiv:1705.03869 [gr-qc]} \BibitemShut
  {NoStop}%
\bibitem [{\citenamefont {Krishnendu}\ \emph
  {et~al.}(2019{\natexlab{a}})\citenamefont {Krishnendu}, \citenamefont
  {Mishra},\ and\ \citenamefont {Arun}}]{Krishnendu:2018nqa}%
  \BibitemOpen
  \bibfield  {author} {\bibinfo {author} {\bibfnamefont {N.~V.}\ \bibnamefont
  {Krishnendu}}, \bibinfo {author} {\bibfnamefont {C.~K.}\ \bibnamefont
  {Mishra}}, \ and\ \bibinfo {author} {\bibfnamefont {K.~G.}\ \bibnamefont
  {Arun}},\ }\href {\doibase 10.1103/PhysRevD.99.064008} {\bibfield  {journal}
  {\bibinfo  {journal} {Phys. Rev.}\ }\textbf {\bibinfo {volume} {D99}},\
  \bibinfo {pages} {064008} (\bibinfo {year} {2019}{\natexlab{a}})},\ \Eprint
  {http://arxiv.org/abs/1811.00317} {arXiv:1811.00317 [gr-qc]} \BibitemShut
  {NoStop}%
\bibitem [{\citenamefont {Krishnendu}\ and\ \citenamefont
  {Yelikar}(2019)}]{Krishnendu:2019ebd}%
  \BibitemOpen
  \bibfield  {author} {\bibinfo {author} {\bibfnamefont {N.~V.}\ \bibnamefont
  {Krishnendu}}\ and\ \bibinfo {author} {\bibfnamefont {A.~B.}\ \bibnamefont
  {Yelikar}},\ }\href@noop {} {\  (\bibinfo {year} {2019})},\ \Eprint
  {http://arxiv.org/abs/1904.12712} {arXiv:1904.12712 [gr-qc]} \BibitemShut
  {NoStop}%
\bibitem [{\citenamefont {Boh\'e}\ \emph {et~al.}(2015)\citenamefont {Boh\'e},
  \citenamefont {Faye}, \citenamefont {Marsat},\ and\ \citenamefont
  {Porter}}]{Bohe:2015ana}%
  \BibitemOpen
  \bibfield  {author} {\bibinfo {author} {\bibfnamefont {A.}~\bibnamefont
  {Boh\'e}}, \bibinfo {author} {\bibfnamefont {G.}~\bibnamefont {Faye}},
  \bibinfo {author} {\bibfnamefont {S.}~\bibnamefont {Marsat}}, \ and\ \bibinfo
  {author} {\bibfnamefont {E.~K.}\ \bibnamefont {Porter}},\ }\href {\doibase
  10.1088/0264-9381/32/19/195010} {\bibfield  {journal} {\bibinfo  {journal}
  {Class. Quant. Grav.}\ }\textbf {\bibinfo {volume} {32}},\ \bibinfo {pages}
  {195010} (\bibinfo {year} {2015})},\ \Eprint
  {http://arxiv.org/abs/1501.01529} {arXiv:1501.01529 [gr-qc]} \BibitemShut
  {NoStop}%
\bibitem [{\citenamefont {Ryan}(1997)}]{Ryan:1996nk}%
  \BibitemOpen
  \bibfield  {author} {\bibinfo {author} {\bibfnamefont {F.~D.}\ \bibnamefont
  {Ryan}},\ }\href {\doibase 10.1103/PhysRevD.55.6081} {\bibfield  {journal}
  {\bibinfo  {journal} {Phys. Rev. D}\ }\textbf {\bibinfo {volume} {55}},\
  \bibinfo {pages} {6081} (\bibinfo {year} {1997})}\BibitemShut {NoStop}%
\bibitem [{\citenamefont {Krishnendu}\ \emph
  {et~al.}(2019{\natexlab{b}})\citenamefont {Krishnendu}, \citenamefont
  {Saleem}, \citenamefont {Samajdar}, \citenamefont {Arun}, \citenamefont
  {Del~Pozzo},\ and\ \citenamefont {Mishra}}]{Krishnendu:2019tjp}%
  \BibitemOpen
  \bibfield  {author} {\bibinfo {author} {\bibfnamefont {N.~V.}\ \bibnamefont
  {Krishnendu}}, \bibinfo {author} {\bibfnamefont {M.}~\bibnamefont {Saleem}},
  \bibinfo {author} {\bibfnamefont {A.}~\bibnamefont {Samajdar}}, \bibinfo
  {author} {\bibfnamefont {K.~G.}\ \bibnamefont {Arun}}, \bibinfo {author}
  {\bibfnamefont {W.}~\bibnamefont {Del~Pozzo}}, \ and\ \bibinfo {author}
  {\bibfnamefont {C.~K.}\ \bibnamefont {Mishra}},\ }\href {\doibase
  10.1103/PhysRevD.100.104019} {\bibfield  {journal} {\bibinfo  {journal}
  {Phys. Rev. D}\ }\textbf {\bibinfo {volume} {100}},\ \bibinfo {pages}
  {104019} (\bibinfo {year} {2019}{\natexlab{b}})},\ \Eprint
  {http://arxiv.org/abs/1908.02247} {arXiv:1908.02247 [gr-qc]} \BibitemShut
  {NoStop}%
\bibitem [{\citenamefont {Narikawa}\ \emph {et~al.}(2021)\citenamefont
  {Narikawa}, \citenamefont {Uchikata},\ and\ \citenamefont
  {Tanaka}}]{Narikawa:2021pak}%
  \BibitemOpen
  \bibfield  {author} {\bibinfo {author} {\bibfnamefont {T.}~\bibnamefont
  {Narikawa}}, \bibinfo {author} {\bibfnamefont {N.}~\bibnamefont {Uchikata}},
  \ and\ \bibinfo {author} {\bibfnamefont {T.}~\bibnamefont {Tanaka}},\
  }\href@noop {} {\  (\bibinfo {year} {2021})},\ \Eprint
  {http://arxiv.org/abs/2106.09193} {arXiv:2106.09193 [gr-qc]} \BibitemShut
  {NoStop}%
\bibitem [{\citenamefont {Allen}\ \emph {et~al.}(2012)\citenamefont {Allen},
  \citenamefont {Anderson}, \citenamefont {Brady}, \citenamefont {Brown},\ and\
  \citenamefont {Creighton}}]{Allen:2005fk}%
  \BibitemOpen
  \bibfield  {author} {\bibinfo {author} {\bibfnamefont {B.}~\bibnamefont
  {Allen}}, \bibinfo {author} {\bibfnamefont {W.~G.}\ \bibnamefont {Anderson}},
  \bibinfo {author} {\bibfnamefont {P.~R.}\ \bibnamefont {Brady}}, \bibinfo
  {author} {\bibfnamefont {D.~A.}\ \bibnamefont {Brown}}, \ and\ \bibinfo
  {author} {\bibfnamefont {J.~D.~E.}\ \bibnamefont {Creighton}},\ }\href
  {\doibase 10.1103/PhysRevD.85.122006} {\bibfield  {journal} {\bibinfo
  {journal} {Phys. Rev. D}\ }\textbf {\bibinfo {volume} {85}},\ \bibinfo
  {pages} {122006} (\bibinfo {year} {2012})},\ \Eprint
  {http://arxiv.org/abs/gr-qc/0509116} {arXiv:gr-qc/0509116} \BibitemShut
  {NoStop}%
\bibitem [{\citenamefont {Veitch}\ \emph {et~al.}(2015)\citenamefont {Veitch}
  \emph {et~al.}}]{Veitch:2014wba}%
  \BibitemOpen
  \bibfield  {author} {\bibinfo {author} {\bibfnamefont {J.}~\bibnamefont
  {Veitch}} \emph {et~al.},\ }\href {\doibase 10.1103/PhysRevD.91.042003}
  {\bibfield  {journal} {\bibinfo  {journal} {Phys. Rev.}\ }\textbf {\bibinfo
  {volume} {D91}},\ \bibinfo {pages} {042003} (\bibinfo {year} {2015})},\
  \Eprint {http://arxiv.org/abs/1409.7215} {arXiv:1409.7215 [gr-qc]}
  \BibitemShut {NoStop}%
\bibitem [{\citenamefont {{Speagle}}(2020)}]{2020MNRAS.tmp..280S}%
  \BibitemOpen
  \bibfield  {author} {\bibinfo {author} {\bibfnamefont {J.~S.}\ \bibnamefont
  {{Speagle}}},\ }\href {\doibase 10.1093/mnras/staa278} {\bibfield  {journal}
  {\bibinfo  {journal} {\mnras}\ } (\bibinfo {year} {2020}),\
  10.1093/mnras/staa278},\ \Eprint {http://arxiv.org/abs/1904.02180}
  {arXiv:1904.02180 [astro-ph.IM]} \BibitemShut {NoStop}%
\bibitem [{\citenamefont {Ashton}\ \emph {et~al.}(2019)\citenamefont {Ashton}
  \emph {et~al.}}]{Ashton:2018jfp}%
  \BibitemOpen
  \bibfield  {author} {\bibinfo {author} {\bibfnamefont {G.}~\bibnamefont
  {Ashton}} \emph {et~al.},\ }\href {\doibase 10.3847/1538-4365/ab06fc}
  {\bibfield  {journal} {\bibinfo  {journal} {Astrophys. J. Suppl.}\ }\textbf
  {\bibinfo {volume} {241}},\ \bibinfo {pages} {27} (\bibinfo {year} {2019})},\
  \Eprint {http://arxiv.org/abs/1811.02042} {arXiv:1811.02042 [astro-ph.IM]}
  \BibitemShut {NoStop}%
\bibitem [{\citenamefont {Romero-Shaw}\ \emph {et~al.}(2020)\citenamefont
  {Romero-Shaw} \emph {et~al.}}]{Romero-Shaw:2020owr}%
  \BibitemOpen
  \bibfield  {author} {\bibinfo {author} {\bibfnamefont {I.~M.}\ \bibnamefont
  {Romero-Shaw}} \emph {et~al.},\ }\href {\doibase 10.1093/mnras/staa2850}
  {\bibfield  {journal} {\bibinfo  {journal} {Mon. Not. Roy. Astron. Soc.}\
  }\textbf {\bibinfo {volume} {499}},\ \bibinfo {pages} {3295} (\bibinfo {year}
  {2020})},\ \Eprint {http://arxiv.org/abs/2006.00714} {arXiv:2006.00714
  [astro-ph.IM]} \BibitemShut {NoStop}%
\bibitem [{\citenamefont {Farr}()}]{time-phase-marginalization}%
  \BibitemOpen
  \bibfield  {author} {\bibinfo {author} {\bibfnamefont {W.~M.}\ \bibnamefont
  {Farr}},\ }\href@noop {} {\enquote {\bibinfo {title} {{Marginalisation of the
  Time Parameter in Gravitational Wave Parameter Estimation.}
  \url{https://dcc.ligo.org/T1400460-v2/public}},}\ }\BibitemShut {NoStop}%
\bibitem [{\citenamefont {Thrane}\ and\ \citenamefont
  {Talbot}(2019)}]{Thrane:2018qnx}%
  \BibitemOpen
  \bibfield  {author} {\bibinfo {author} {\bibfnamefont {E.}~\bibnamefont
  {Thrane}}\ and\ \bibinfo {author} {\bibfnamefont {C.}~\bibnamefont
  {Talbot}},\ }\href {\doibase 10.1017/pasa.2019.2} {\bibfield  {journal}
  {\bibinfo  {journal} {Publ. Astron. Soc. Austral.}\ }\textbf {\bibinfo
  {volume} {36}},\ \bibinfo {pages} {e010} (\bibinfo {year} {2019})},\ \Eprint
  {http://arxiv.org/abs/1809.02293} {arXiv:1809.02293 [astro-ph.IM]}
  \BibitemShut {NoStop}%
\bibitem [{\citenamefont {Singer}\ and\ \citenamefont
  {Price}(2016)}]{Singer:2015ema}%
  \BibitemOpen
  \bibfield  {author} {\bibinfo {author} {\bibfnamefont {L.~P.}\ \bibnamefont
  {Singer}}\ and\ \bibinfo {author} {\bibfnamefont {L.~R.}\ \bibnamefont
  {Price}},\ }\href {\doibase 10.1103/PhysRevD.93.024013} {\bibfield  {journal}
  {\bibinfo  {journal} {Phys. Rev. D}\ }\textbf {\bibinfo {volume} {93}},\
  \bibinfo {pages} {024013} (\bibinfo {year} {2016})},\ \Eprint
  {http://arxiv.org/abs/1508.03634} {arXiv:1508.03634 [gr-qc]} \BibitemShut
  {NoStop}%
\bibitem [{\citenamefont {Singer}\ \emph {et~al.}(2016)\citenamefont {Singer}
  \emph {et~al.}}]{Singer:2016eax}%
  \BibitemOpen
  \bibfield  {author} {\bibinfo {author} {\bibfnamefont {L.~P.}\ \bibnamefont
  {Singer}} \emph {et~al.},\ }\href {\doibase 10.3847/2041-8205/829/1/L15}
  {\bibfield  {journal} {\bibinfo  {journal} {Astrophys. J. Lett.}\ }\textbf
  {\bibinfo {volume} {829}},\ \bibinfo {pages} {L15} (\bibinfo {year}
  {2016})},\ \Eprint {http://arxiv.org/abs/1603.07333} {arXiv:1603.07333
  [astro-ph.HE]} \BibitemShut {NoStop}%
\bibitem [{\citenamefont {Kidder}\ \emph {et~al.}(1992)\citenamefont {Kidder},
  \citenamefont {Will},\ and\ \citenamefont {Wiseman}}]{Kidder:1992zz}%
  \BibitemOpen
  \bibfield  {author} {\bibinfo {author} {\bibfnamefont {L.~E.}\ \bibnamefont
  {Kidder}}, \bibinfo {author} {\bibfnamefont {C.~M.}\ \bibnamefont {Will}}, \
  and\ \bibinfo {author} {\bibfnamefont {A.~G.}\ \bibnamefont {Wiseman}},\
  }\href {\doibase 10.1088/0264-9381/9/9/004} {\bibfield  {journal} {\bibinfo
  {journal} {Class. Quant. Grav.}\ }\textbf {\bibinfo {volume} {9}},\ \bibinfo
  {pages} {L125} (\bibinfo {year} {1992})}\BibitemShut {NoStop}%
\bibitem [{\citenamefont {Blanchet}(2002)}]{Blanchet:2001id}%
  \BibitemOpen
  \bibfield  {author} {\bibinfo {author} {\bibfnamefont {L.}~\bibnamefont
  {Blanchet}},\ }\href {\doibase 10.1103/PhysRevD.65.124009} {\bibfield
  {journal} {\bibinfo  {journal} {Phys. Rev.}\ }\textbf {\bibinfo {volume}
  {D65}},\ \bibinfo {pages} {124009} (\bibinfo {year} {2002})},\ \Eprint
  {http://arxiv.org/abs/gr-qc/0112056} {arXiv:gr-qc/0112056 [gr-qc]}
  \BibitemShut {NoStop}%
\bibitem [{\citenamefont {Helstrom}(1995)}]{Helstrom:1994}%
  \BibitemOpen
  \bibfield  {author} {\bibinfo {author} {\bibfnamefont {C.~W.}\ \bibnamefont
  {Helstrom}},\ }\href@noop {} {\emph {\bibinfo {title} {Elements of Signal
  Detection and Estimation}}}\ (\bibinfo  {publisher} {Prentice-Hall, Inc.},\
  \bibinfo {address} {Upper Saddle River, NJ, USA},\ \bibinfo {year}
  {1995})\BibitemShut {NoStop}%
\bibitem [{\citenamefont {Akmal}\ \emph {et~al.}(1998)\citenamefont {Akmal},
  \citenamefont {Pandharipande},\ and\ \citenamefont
  {Ravenhall}}]{Akmal:1998cf}%
  \BibitemOpen
  \bibfield  {author} {\bibinfo {author} {\bibfnamefont {A.}~\bibnamefont
  {Akmal}}, \bibinfo {author} {\bibfnamefont {V.~R.}\ \bibnamefont
  {Pandharipande}}, \ and\ \bibinfo {author} {\bibfnamefont {D.~G.}\
  \bibnamefont {Ravenhall}},\ }\href {\doibase 10.1103/PhysRevC.58.1804}
  {\bibfield  {journal} {\bibinfo  {journal} {Phys. Rev. C}\ }\textbf {\bibinfo
  {volume} {58}},\ \bibinfo {pages} {1804} (\bibinfo {year} {1998})},\ \Eprint
  {http://arxiv.org/abs/nucl-th/9804027} {arXiv:nucl-th/9804027} \BibitemShut
  {NoStop}%
\bibitem [{\citenamefont {Pappas}\ and\ \citenamefont
  {Apostolatos}(2012)}]{Pappas:2012qg}%
  \BibitemOpen
  \bibfield  {author} {\bibinfo {author} {\bibfnamefont {G.}~\bibnamefont
  {Pappas}}\ and\ \bibinfo {author} {\bibfnamefont {T.~A.}\ \bibnamefont
  {Apostolatos}},\ }\href@noop {} {\  (\bibinfo {year} {2012})},\ \Eprint
  {http://arxiv.org/abs/1211.6299} {arXiv:1211.6299 [gr-qc]} \BibitemShut
  {NoStop}%
\bibitem [{\citenamefont {Shoemaker}()}]{design-sensitivity}%
  \BibitemOpen
  \bibfield  {author} {\bibinfo {author} {\bibfnamefont {D.}~\bibnamefont
  {Shoemaker}},\ }\href@noop {} {\enquote {\bibinfo {title} {{Advanced LIGO
  anticipated sensitivity curves.}
  \url{https://dcc.ligo.org/LIGO-T0900288/public}},}\ }\BibitemShut {NoStop}%
\bibitem [{\citenamefont {Pearson}(1992)}]{Pearson1992}%
  \BibitemOpen
  \bibfield  {author} {\bibinfo {author} {\bibfnamefont {K.}~\bibnamefont
  {Pearson}},\ }\enquote {\bibinfo {title} {On the criterion that a given
  system of deviations from the probable in the case of a correlated system of
  variables is such that it can be reasonably supposed to have arisen from
  random sampling},}\ in\ \href {\doibase 10.1007/978-1-4612-4380-9_2} {\emph
  {\bibinfo {booktitle} {Breakthroughs in Statistics: Methodology and
  Distribution}}},\ \bibinfo {editor} {edited by\ \bibinfo {editor}
  {\bibfnamefont {S.}~\bibnamefont {Kotz}}\ and\ \bibinfo {editor}
  {\bibfnamefont {N.~L.}\ \bibnamefont {Johnson}}}\ (\bibinfo  {publisher}
  {Springer New York},\ \bibinfo {address} {New York, NY},\ \bibinfo {year}
  {1992})\ pp.\ \bibinfo {pages} {11--28}\BibitemShut {NoStop}%
\bibitem [{\citenamefont {Neyman}\ and\ \citenamefont
  {Pearson}(1992)}]{Neyman1992}%
  \BibitemOpen
  \bibfield  {author} {\bibinfo {author} {\bibfnamefont {J.}~\bibnamefont
  {Neyman}}\ and\ \bibinfo {author} {\bibfnamefont {E.~S.}\ \bibnamefont
  {Pearson}},\ }\enquote {\bibinfo {title} {On the problem of the most
  efficient tests of statistical hypotheses},}\ in\ \href {\doibase
  10.1007/978-1-4612-0919-5_6} {\emph {\bibinfo {booktitle} {Breakthroughs in
  Statistics: Foundations and Basic Theory}}},\ \bibinfo {editor} {edited by\
  \bibinfo {editor} {\bibfnamefont {S.}~\bibnamefont {Kotz}}\ and\ \bibinfo
  {editor} {\bibfnamefont {N.~L.}\ \bibnamefont {Johnson}}}\ (\bibinfo
  {publisher} {Springer New York},\ \bibinfo {address} {New York, NY},\
  \bibinfo {year} {1992})\ pp.\ \bibinfo {pages} {73--108}\BibitemShut
  {NoStop}%
\end{thebibliography}%
\end{document}